\documentclass[letterpaper,nofootinbib,onecolumn,eqsecnum,superscriptaddress,floatfix,prd
]{revtex4}

\usepackage[linktocpage]{hyperref}
\usepackage{xcolor}
\usepackage{tabularx}
\usepackage{soul}
\usepackage{amstext}
\usepackage{fancyhdr}
\usepackage{verbatim}
\usepackage{mathrsfs}
\usepackage{amsfonts}
\usepackage{hyperref}
\usepackage{amssymb}
\usepackage{amsmath}
\usepackage{latexsym}
\usepackage{graphicx}
\usepackage{color}
\usepackage{float} 
\usepackage{ulem}

\hypersetup{
    colorlinks,
    linkcolor={blue!60!black},
    citecolor={blue!80!black},
    urlcolor={blue!20!black}
}

\begin{document}


\title{On the detectability of gravitational waves emitted from head-on collisions of $\ell$-boson stars} 

\author{Mariana Lira}
\email[]{lira@ciencias.unam.mx} 
\affiliation{Instituto de Ciencias Nucleares, Universidad Nacional
  Aut\'onoma de M\'exico, Circuito Exterior C.U., A.P. 70-543,
  Coyoac\'an, M\'exico 04510, CdMx, M\'exico}

\author{Laura O. Villegas}
\email[]{laura.villegas8344@alumnos.udg.mx} 
\affiliation{Departamento de F\'isica,
Centro Universitario de Ciencias Exactas e Ingenier\'ia, Universidad de Guadalajara\\
Av. Revoluci\'on 1500, Colonia Ol\'impica C.P. 44430, Guadalajara, Jalisco, M\'exico}

\author{Javier M. Antelis} 
\email[]{mauricio.antelis@tec.mx} 
\affiliation{Tecnologico de Monterrey, Escuela de Ingenier\'ia y Ciencias\\
Av. Eugenio Garza Sada 2501 Sur, Colonia Tecnológico \\
Monterrey, N.L., 64849, México}

\author{V\'ictor Jaramillo }
\email[]{jaramillo@ustc.edu.cn} 
\affiliation{Department of Astronomy, University of Science and Technology of China, Hefei, Anhui 230026, China}
\affiliation{Department of Modern Physics, University of Science and Technology of China, Hefei, Anhui 230026, China}

\author{Claudia Moreno} 
\email[]{claudia.moreno@academico.udg.mx} 
\affiliation{Departamento de F\'isica,
Centro Universitario de Ciencias Exactas e Ingenier\'ia, Universidad de Guadalajara\\
Av. Revoluci\'on 1500, Colonia Ol\'impica C.P. 44430, Guadalajara, Jalisco, M\'exico}
  
\author{Dar\'io N\'u\~nez }
\email[]{nunez@nucleares.unam.mx} 
\affiliation{Instituto de Ciencias Nucleares, Universidad Nacional
  Aut\'onoma de M\'exico, Circuito Exterior C.U., A.P. 70-543,
  Coyoac\'an, M\'exico 04510, CdMx, M\'exico}
\affiliation{Departamento de Matemática da Universidade de Aveiro and Centre for Research and Development in Mathematics and Applications (CIDMA), Campus de Santiago, 3810-183 Aveiro, Portugal.}


\date{November 30, 2024}

%
\bigskip

\begin{abstract}
In this work, we investigate head-on collisions of $\ell$-boson stars, potential candidates for dark matter compact objects. We begin with a review of the general properties and features of these stars, leveraging results from prior studies to analyze the gravitational wave signals generated by such collisions. Considering a maximum distance of 100 Mpc for potential events, we identify the range of masses and scalar field frequencies for these stars that would render the gravitational waves detectable by current gravitational wave observatories. Additionally, we process the resulting signals to generate simulated observatory images, highlighting their similarities and differences compared to those produced by black hole collisions.
\end{abstract}

\maketitle


\tableofcontents

\section{Introduction}

As we advance further into the gravitational wave detection era, the prospect of identifying waveforms beyond those typically associated with black hole collisions becomes increasingly intriguing. For instance, events such as core-collapse supernovae can also generate gravitational waves with intensities detectable by current laser interferometry-based observatories. Significant efforts have been dedicated to the numerical simulation of such events \cite{KoSaKa06,AbPaRa22,Andresen2019,Radice2019,Morozova2018,Mezzacappa:2020lsn,Muller:2012,OConnor:2010moj}. A detailed account of the mechanisms behind supernova gravitational wave production, their propagation, and the potential for detection is provided in \cite{Szczepanczyk:2023ihe}.

The LIGO-VIRGO-KAGRA~\cite{TheLIGOScientific:2014jea, TheVirgo:2014hva, Aso:2013eba} collaboration has finished three observing runs and released the results in the third Gravitational-Waves Transient Catalog (GWTC-3) \cite{LIGOScientific:2021usb,LIGOScientific:2021qlt,LIGOScientific:2021djp}. In total, the collaboration has observed more than 
90 events since 2015, when it began operating. All these events are associated with gravitational waves emitted during the collision of black holes and neutron stars binaries systems. Run O4 started in May 2023 and it is expected that the results of the fourth observation period contain information that allows for statistics on compact binaries, in addition to the awaited supernova gravitational waves \cite{Maggiore:2007ulw}. During the past two years, 81 highly likely gravitational event candidates have been identified in O4 \cite{LIGO:2024kkz}. 

New astrophysical objects that have not yet been experimentally detected could be discovered, such as those composed of dark matter, known as dark stars \cite{Spolyar:2007qv,Freese:2015mta}. An interesting work that analyzes the emission of gravitational waves in alternative theories of compact binary systems can be reviewed in \cite{Casallas-Lagos:2022nwd}. Actually, in \cite{Bustillo:2020syj}, it was discussed the 
fact that the gravitational waves could present a degeneracy. Indeed, the authors focused on the gravitational wave GW190521 \cite{LIGOScientific:2020iuh} detected, and showed that the signal could not only be explained as generated by the collision of the $150\,M_\odot$ binary black hole merger, but also by the collision of Proca stars, which are self gravitating configurations composed of Proca fields. Moreover, the authors claim that the former case is even more plausible. In any case, the clear fact is that there are different possible sources of a gravitational signal. 
On the other side of the hypothesis of the existence of several sources of gravitational waves, one can explore the possibility of detecting signals of the collision of objects that are appreciably distinguishable from black holes and compact neutron stars. Certainly, scalar and vector field solitons with high compactness have subtle impact on the characteristic gravitational wave signal compared to the black hole paradigm. In contrast, systems with low compactness produce distinctive signatures. A free parameter in this model, related to the boson mass $\mu$, allows for {\bf rescaling} the gravitational signal in these systems. Like black holes in general relativity, these signals might fall within the detection range of current interferometers.

One of these possible sources of gravitational waves are those generated by the collision of horizonless exotic compact objects; the Proca stars mentioned above is one such example. One simpler model is the collision of boson stars, it is considered simpler because, in the absence of interactions with other forms of matter, the dynamics are governed solely by a wave equation (the classical Klein-Gordon equation) and the gravitational interactions \cite{Palenzuela:2006wp}. This model is particularly interesting because the scalar field is a strong candidate for describing dark matter in the Universe \cite{Matos:1998vk, Matos:1999et, Suarez:2013iw, Schive:2014dra, Hui:2016ltb}. The compact objects formed by this field, such as boson stars and related structures, make the detection of gravitational waves from their collisions significant—not only in itself but also as potential evidence for compact objects composed of dark matter.

Indeed, dark matter presence has the indirect evidence on the dynamics of observable matter but, if there are compact objects made of dark matter, {\it e.g}, boson stars, being gravitational objects, a possible way to directly probe the presence of boson stars in nature,
and possible of dark matter itself, is by the emission of gravitational waves during a collision of two of these objects. The generation of such gravitational waves has been studied in \cite{Palenzuela:2006wp}, where the authors used numerical relativity techniques to evolve in axial symmetry and full general relativity the head-on collision of two of the simplest model of boson stars, corresponding to the static spherically symmetric and free-field family of solutions. 

The self-gravitating solutions of complex, massive scalar fields, commonly referred to as mini-boson stars, were first discovered over 50 years ago \cite{Kaup:1968zz}. Since then, numerous generalizations have been explored (see \cite{Liebling:2012fv,Shnir:2022lba} for comprehensive reviews). One such generalization is the $\ell$-boson stars \cite{Alcubierre:2018ahf,Alcubierre:2021psa}, which consist of an odd number ($2\ell+1$) of non-spherical scalar fields. The case $\ell=0$ corresponds to the standard mini-boson star. These gravitational soliton configurations are constructed such that the energy-momentum tensor, and thus the spacetime metric of the equilibrium solutions, remains spherically symmetric. Like mini-boson stars, $\ell$-boson stars possess the critical property of stability \cite{Alcubierre:2021mvs,Jaramillo:2020rsv}. In \cite{Jaramillo:2022zwg}, head-on collisions of $\ell=0$ and $\ell=1$ boson stars were studied using fully non-linear numerical relativity simulations in three dimensions, with a detailed analysis of the resulting gravitational wave emission.

In \cite{Jaramillo:2022zwg}, gravitational waves generated by $\ell$-boson stars were clearly linked to the compactness of the colliding stars. It is important to note that these objects do not collide like fluid stars; instead, they pass through each other, producing interference patterns through their gravitational interaction. Depending on the initial conditions and the gravitational well, $\ell$-boson stars can pass through one another, exhibit interference, and then continue their trajectories.
Although no direct collision occurs, the gravitational and scalar interactions in more compact scenarios can be so intense that the effect on the surrounding geometry becomes comparable to that of black hole collisions. In fact, the gravitational wave profiles generated in some of these encounters closely resemble those produced by black holes, with a sharp peak and a ringdown phase. In certain cases, the final state after the collision of highly compact $\ell$-boson stars is a black hole. On the other hand, when no black hole is formed and a single remnant object is obtained, it has been observed that the object takes a significant amount of time to stabilize. This seems to be a characteristic feature of $\ell$-boson stars, as their dissipation mechanisms are quite slow. In \cite{Jaramillo:2022zwg}, it was noted that after the plunge, a bounded object remained within a specific region, but showed no clear signs of settling. Similarly, in \cite{CalderonBustillo:2020fyi}, gravitational waves generated after the merger of the above mentioned Proca stars binary system (which are similar to boson stars but consist of a massive vector field, see \cite{Brito:2015pxa}) were analyzed. The ability of these plunges to produce gravitational signals similar to those of black holes is crucial, as it suggests that boson and Proca stars can act as black hole mimickers. However, in this work, we will explore a different approach.

In this work, we investigate gravitational wave profiles that differ from the typical signatures of black hole collisions. First, we examine cases where head-on collisions of $\ell=0$ or $\ell=1$ boson stars could produce gravitational waves with features detectable by gravitational observatories. We set parameters for the scalar field mass, $\mu$, as well as the compactness, masses, sizes, and distances, so that the resulting gravitational waves passing Earth would have a measurable impact on observational instruments. Specifically, we aim for signals that could be captured in the data from the upcoming O5 LIGO run.

This paper is organized as follows. In Section \ref{head on collision} we present a detailed summary of the derivation of the mini-boson stars, ($\ell=0$ boson stars in our notation), as self gravitating configurations, describing their main properties; next proceed to discuss the general $\ell$ boson stars, and finally describe the head-on collision of several $\ell$ boson stars configurations. In Section \ref{Sec:results}, we study the properties of gravitational wave generated, by means of the polarization modes, and determine the conditions such that they could be detected by the present day gravitational observatories, as well as define the images that would be seen. Finally, in Section \ref{conclusions} we present some final remarks and point out some possible lines for subsequent works.

\section{\texorpdfstring{$\ell$}{l}-boson stars and their head-on collision} 
\label{head on collision}

\subsection{\texorpdfstring{$\ell$}{l} boson star}

The scalar field satisfies the Klein-Gordon equation, which is the wave equation that describes the relativistic generalization of 
the Schr\"odinger equation. As such, the scalar field inherits the fuzzy concept of being related to a distribution of probability. However, in the context of the Schr\"odinger-Poisson system, the nonrelativistic limit of the Einstein-Klein-Gordon system, it is possible to treat the wave function as a mass probability that generates a gravitational field, which being the source of the Poisson equation. In the relativistic description, it can determine the geometry of the space time by means of a well-defined stress energy tensor and thus being the source in the Einstein equations. In this way, it can be analyzed in a cosmological context with Friedman like geometries or in a local one, defining compact, self gravitating objects. The scalar field can be a complex function, $\Phi=\Phi(x^\mu)$ and the corresponding field equation and stress energy tensor can be derived from a Lagrangian:
\begin{equation} \label{eq:Lag}
{\cal L}=-\frac{k}{2}\,\left(\nabla^\mu\,\Phi\,\nabla_\mu\,\Phi^* + V\left(|\Phi|\right)\right),    
\end{equation}
with $k$ a constant related to the units used, and $|\Phi|=\Phi\,\Phi^*$ is the norm of the scalar field.

A remark on such a constant $k$ included in the Lagrangian, Eq.~(\ref{eq:Lag}), is needed to have a clearer understanding of the scalar field properties. Such constant, $k$, passes to the stress energy tensor, Eq.~(\ref{Tmunu_escc}) {\it via} variational principle. On the one hand, we have that the Einstein equations, Eqs.~(\ref{eq:Eins}), demand that the stress energy tensor has units of {\it pressure}, $\frac{{\rm M}}{{\rm L}{\rm T}^2}$, standing for mass, ${\rm M}$, length, ${\rm L}$ and time, ${\rm T}$. On the other hand, from the stress energy tensor, Eq.~(\ref{Tmunu_escc}), the presence of the nabla operators introduces units of inverse length square, ${\rm L}^2$. From this reasoning, it can be seen that essentially $k\,[\Phi]^2$, with $[\Phi]$ standinng for the units of the scalar field, can be taken as the inverse of the constants in the Einstein equations $k=\frac{c^4}{8\,\pi\,G}$, we include the factor $\frac{1}{8\,\pi}$ for completeness. One can considered the scalar field to be unitless $[\Phi]=1$ (consistent with the notion of probability function), and then the constant $k$ has units of force, $\frac{{\rm M}\,{\rm L}}{{\rm T}^2}$. Alternatively, the constant can be absorbed in the scalar field, {\it i. e.} $k=1$, and then the scalar field has units of $\frac{c^2}{\sqrt{G}}$ which in the unit system used in particle physics, $c=\hbar=1$, the gravitational constant $G$ is identified with the inverse of the square of Planck's mass, so that in these units, the scalar field is given in this choice with units of Planck's mass and the constant in front of the stress energy tensor, Eq.~(\ref{Tmunu_escc}) is taken as one. Interestingly, the units of the scalar field, unitless, Planckaninan mass, or even other combinations, do not affect the physical quantities, as long as they are taken consistently, in the stress energy tensor or in the scalar field. In this work, we use $k=1$, which is the most common usage.

The stress energy tensor of the scalar field, obtained from the Lagrangian \eqref{eq:Lag} is (recall $k$ will be taken as equal to one),
\begin{equation}
T_{\mu\nu} = \frac{k}{2}\,\left( \nabla_{\mu}\Phi\,\nabla_\nu\,\Phi^* + \nabla_{\nu}\Phi\,\nabla_{\mu}\,\Phi^*  - g_{\mu\nu} \left(g^{\alpha\beta}\nabla_{\alpha}\Phi\,\nabla_{\beta} \Phi^* + V\left(|\Phi|^2\right)\right) \right) \, .  \label{Tmunu_escc}   
\end{equation}

The above mentioned Klein-Gordon equation is the equation of motion obtained from the variation of the Lagrangian with respect to the (conjugate) of the complex scalar field
\begin{equation}
 g^{\alpha\beta}\,\nabla_\alpha\,\nabla_\beta\,\Phi - \frac{\partial V}{\partial \Phi^*}=0,   
\end{equation}
this expression is essentially a wave equation, and describes the dynamics of the scalar field in a given space time determined by the metric tensor $g_{\mu\nu}$.
When solved together, the Einstein equations with the stress energy tensor given by Eq.~(\ref{Tmunu_escc}), a self-gravitating structure composed of a scalar field can be derived.

In the case where one consider the simple case of a static spherically symmetric spacetime, it is usuall to give the anzatz for the scalar field as a plane wave with frequency $\omega$:
\begin{equation}
\Phi=\phi(r)\,e^{i\,\omega\,t},    
\end{equation}
and considered the scalar potential $V$ as
\begin{equation}
V\left(|\Phi|^2\right)=\mu^2\,\phi^2,  \label{eq:Vesc}   
\end{equation}
with $\mu$ a parameter with units of $(\mathrm{length})^{-1}$. The potential can be related to the mass of an associated bosonic particle, $m_\phi$ as $\mu=\frac{m_\phi\,c}{\hbar}$, with $c$ representing the speed of light in vacuum and $\hbar$ representing Planck's constant divided by $2\,\pi$. 

According to the static and spherically symmetric spacetime assumptions of the boson star it is useful to consider the following spacetime line element \cite{Alcubierre:2018ahf}:
\begin{equation}
ds^2 = -\alpha(r)\,c^2\,dt^2 + \frac{dr^2}{1-\frac{2\,M(r)\,G}{c^2\,r}} + r^2 d \Omega^2, \label{eq:le}
\end{equation}
being $\alpha$ and $M$ functions of $r$, and $d\Omega$ is the standard line element in the unit of two-sphere.

The Einstein-Klein-Gordon system is:
\begin{eqnarray}
G_{\alpha\beta}&=&\frac{8\pi G}{c^4} T_{\alpha\beta}, \label{eq:Eins}\\
\square \Phi &=& \mu^2 \Phi, 
\end{eqnarray}
which for the spherically symmetric static space time described by the metric in Eq.~(\ref{eq:le}) leads to a system of three differential equations for the metric functions $\alpha$ and $M$ and for the radial function of the scalar field, $\phi$. The requirement for asymptotic flatness is met under the condition $\omega^2 / c^2 < \mu^2$. This system of equations can be solved numerically to obtain a self-gravitating object composed of a scalar field with mass $m_\phi c^2$ (commonly normalized to 1 by leveraging the spatial rescaling permitted in the Einstein-Klein-Gordon system; see below). The mass of the star decreases as its radius increases, allowing the definition of a characteristic radius, $R_{99}$, which encloses 99\% of the total mass, $M_{99}$. Such an object is referred to as a boson star.

When solving the Einstein-Klein-Gordon equations, the central value of the scalar field is used as an input parameter. A shooting finite differences method \cite{Press:1992zz, Alcubierre:2018ahf} or a spectral method \cite{Grandclement:2007sb, Alcubierre:2022rgp} determines the parameter $\omega$ required for asymptotic flatness, yielding a boson star configuration. By varying the central scalar field value, a family of configurations can be generated, as illustrated in Fig.~\ref{fig:caracolito}. In this figure, the mass $M_{99}$ is plotted against the frequency $\omega$ for each configuration.

\subsection{\texorpdfstring{$\ell$}{l}-boson stars}

As expected, there are several generalizations that can be made to the boson star model \cite{Kaup:1968zz} which is a single field with a linear contribution from the potantial in the Klein-Gordon equation. Actually, they are called mini-boson stars as long as, when they were proposed, the boson particles that were fashion were the axions, whose masses (times the squared speed of light) of the order of $10^{-5}\,{\rm eV}$ and the solf gravitating objects that were determined had masses four orders of magnitude smaller than the sun and small sizes for not very diluted objects, and they were dubbed {\it mini}. Actually, there is a large range for the values of the mass of the bosonic particle, as well as combinations of the mass and compactness that make them not ``mini". In any case, they are simple one-parameter self gravitating spherical objects. A very interesting generalization is to consider scalar potentials $V(|\Phi|^2)$, with more terms than in Eq.~(\ref{eq:Vesc}), including quartic where the coefficient is called the self-interaction term; see \cite{Colpi:1986ye} for a very illuminating introduction to this term and the effects on the final object; or even a sixth order one which has also received more attention \cite{Lynn:1988rb,Macedo:2013jja,Gao:2023gof}.

Another venue for generalization is to leave the spherical symmetry and explore other configurations. It is also natural to consider a generalization that contains several classical scalar fields 
\cite{Alcubierre:2022rgp}, arrange in so a way that the final configuration still is spherical; they were called {\it $\ell$-boson star}. Indeed, for such stars, instead of eq. \eqref{eq:Lag} the following (generic at the beginning) Lagrangian combining an odd number of free scalar fields is considered
\begin{equation} \label{eq:Lagell}
{\cal L}=-\frac{k}{2}\,\sum_{{\rm{m}}=-\ell}^{\ell}\left(\nabla^\mu\,\Phi_{\rm{m}}\,\nabla_\mu\,\Phi_{\rm{m}}^* + V\left(|\Phi_{\rm{m}}|\right)\right),    
\end{equation}
where $\ell$ and m, are the orbital quantum number $-\ell \leq$ m $\leq \ell$.
In \cite{Sanchis-Gual:2021edp}, a detailed analysis is presented on stationary solutions that can be constructed starting from this kind of combinations and the dynamical properties of these configurations. A very remarkable configuration is one in which, for a given value of the parameter $\ell$, the components for all the values of $\rm{m}$ have the same radial profiles but different angular distributions;
by this we mean that associated to each $\Phi_{\rm{m}}$ (given $\ell$, i.e., fixing the number of scalar fields of the theory) there is a spherical harmonic $Y^{\ell,\rm{m}(\theta,\varphi)}=P_\ell(\cos\theta)\,e^{i\,\rm{m}\,\varphi}$ that multiplies it.

In particular, since $\ell$-boson stars are also stationary solutions to the Einstein equations, a harmonic time dependence is also assumed in the scalar field(s). Thus, each of the $2\,\ell + 1$ fields have the form:
\begin{equation}\label{eq:ansatz_ell}
    \Phi_{\rm{m}} (t,r,\vartheta, \varphi ) = e^{-i\omega t} \phi_{\ell}(r)\, Y_{\ell \rm{m}} (\vartheta, \varphi),
\end{equation}
so, the stress energy tensor of the material components given by Eq. \eqref{eq:Lagell}, which is obtained summing up the $2\ell+1$ terms of the form given by the single scalar field stress energy tensor Eq.~(\ref{Tmunu_escc}),
\begin{equation}
T_{\mu\nu}=\frac{k}{2}\,\sum\limits_{\rm{m}} \left(\nabla_\mu\,\Phi_{\rm{m}}\,\nabla_\nu\,{\Phi^*}_{\rm{m}} + \nabla_\nu\,\Phi_{\rm{m}}\,\nabla_\mu\,{\Phi^*}_{\rm{m}}  - g_{\mu\nu}\,\left(g^{\alpha\beta}\,\nabla_\alpha\,\Phi_{\rm{m}}\,\nabla_\beta\,{\Phi^*}_{\rm{m}}+ \mu^2\,\Phi_{\rm{m}}\,{\Phi^*}_{\rm{m}} \right)\right),     
\end{equation}
which generates a spherically symmetric configuration (and independent of coordinate $t$). To see this geometric arrange, one uses the addition theorem of the spherical harmonics:
\begin{equation}
\sum\limits_{\rm{m}=-\ell}^\ell |Y^{\ell \rm{m}}(\vartheta,\varphi)|^2 = \frac{2\ell+1}{4\pi},
\label{Eq:AdditionTheorem}
\end{equation}
and together with the fact that they are eigenfunctions of the angular Laplacian operator:
\begin{equation}
\left(\frac{\partial^2}{\partial\,\theta^2} + \cot\theta\,\frac{\partial}{\partial\,\theta} + \frac{1}{\sin^2\theta}\,\frac{\partial^2}{\partial\,\varphi^2}\right)\, Y^{\ell \rm{m}} = -\ell(\ell+1) Y^{\ell \rm{m}},    
\label{eq:Phi-l}
\end{equation}
it can be proved, see the Appendix in \cite{Alcubierre:2018ahf}, that the angular dependence disappears, the stress energy tensor is spherically symmetric retaining, however, terms including $\ell$ so that the information about the number of scalar fields affect non-trivially to the gravitational and scalar fields since the $\ell$ terms certainly affect the final obtained configurations of equilibrium. This remarkable property was noticed and discussed in \cite{Olabarrieta:2007di}. 

\begin{figure} [H]
   \centering    
   \includegraphics[width=0.45\textwidth]{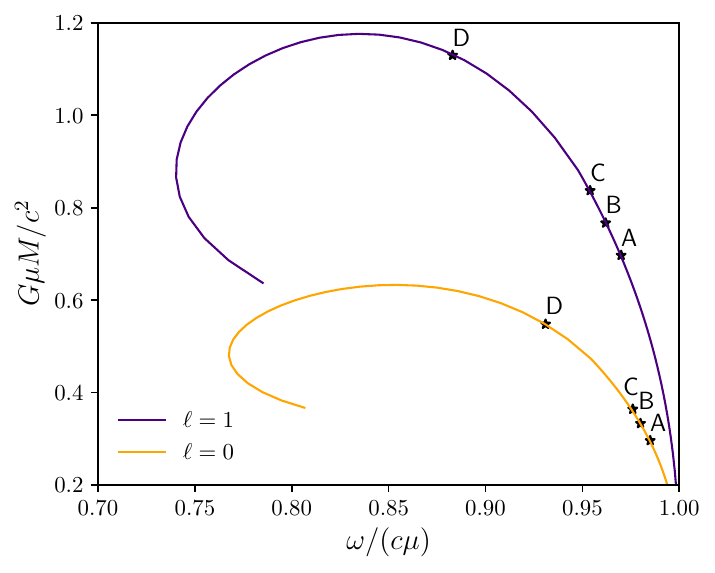}
   \includegraphics[width=0.442\textwidth]{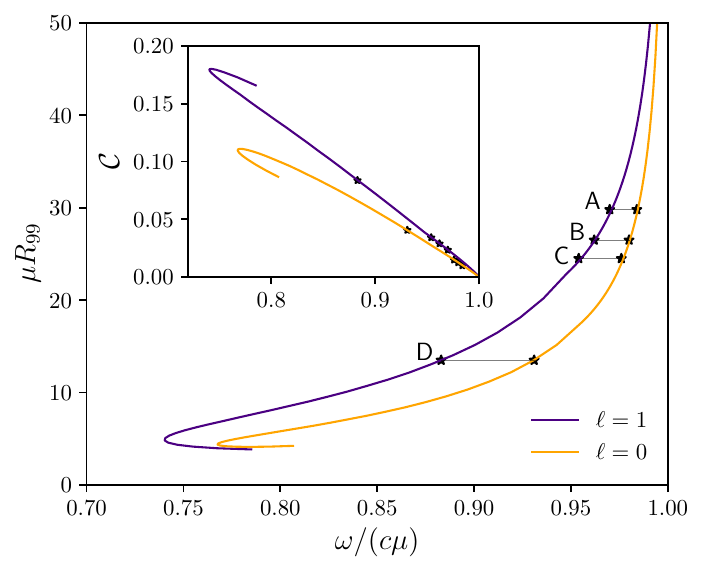}
    \caption{Left: Total mass $M$ of isolated boson stars for the single field $\ell=0$ case and $\ell=1$ versus the frequency oscillation of the scalar field, $\omega$. Right: Radius and compactness (inset) of the same sequences of boson stars. We mark in both plots the cases used for head-on collision in this work, shown in Table \ref{tab:models}.}\label{fig:caracolito}
\end{figure}

The Klein-Gordon equation for the scalar field comprising the $2\,\ell + 1$ scalar fields, given by Eq.~(\ref{eq:Phi-l}), also looses all the angular dependence, but remains a term with the $\ell$ parameter, namely $\ell\,\left(\ell+ 1\right)/r^2$, which is equivalent to the the total angular momentum term which defines the effective potential in Classical Mechanics, as well as the Hydrogen atom problem in Quantum Mechanics. With the scalar field given by the ansatz in Eq. \eqref{eq:ansatz_ell} and the sum in the harmonics we calculate the sources of the Einstein equations and imposing regularity and asymptotic flatness, one obtains again a self gravitating spherically symmetric scalar field configuration, with a radius confining the $99\,\%$ of the mass finite, and an extra parameter: $\ell$. Such configurations were dubbed $\ell$-boson stars \cite{Alcubierre:2018ahf}. 

These objects generalize the usual boson star configuration ($\ell=0$ boson star in this notation). For a given $\ell$, a family of solutions can also be obtained with a snail-like graph in the plot of the mass against the frequency, as the one shown in Fig.~\ref{fig:caracolito} which includes the mini-boson star and the $\ell=1$ case. A remarkable feature of these graphs is that the 
 maximum in the mass separates the stable configurations (to the right) from the unstable ones (to the left of the maximum) \cite{Alcubierre:2021mvs,Jaramillo:2020rsv}. In \cite{Sanchis-Gual:2021edp}, several properties of multiple scalar field configurations were analyzed, and the $\ell$-boson stars were also analyzed, dubbed {\it symmetry-enhanced solutions}; the $\ell$-boson star thoroughly studied in \cite{Alcubierre:2021mvs,Jaramillo:2020rsv}, and it was found the $\ell$-boson stars were stable under spherical perturbations, \cite{Alcubierre:2021mvs} actually probing that for each fixed value $\ell$ of the spherical harmonic function, there exists a family of $\ell$-boson stars which are linearly stable with respect to radial fluctuations, and in \cite{Jaramillo:2020rsv} was seen that the $\ell=1$ boson star, were the branching off towards a larger
family of equilibrium solutions, conjecturing that $1$-boson stars are the enhanced isometry point of a larger family of static (and possibly stationary), non-spherical multi-field self-gravitating solitons. 

Analyzing a semi-classical correspondence, \cite{Alcubierre:2022rgp}, it was seen that these $\ell$-boson star configurations can be compared to distributions with the same quantum number, a sort of $\ell$-condensate. The case for large values, actually extreme values of the parameter $\ell$ was analyzed in \cite{Alcubierre:2021psa} and was seen that large $\ell$ configurations tend to form a shell like structures. Even a similar case using exotic matter was considered in \cite{Carvente:2019gkd} and an $\ell$-worm hole was constructed.

\subsection{Head-on collisions}

As mentioned in the introduction, even though the $\ell$-boson stars do not collide, by means of the combined gravitational interaction, they generate an interference which in turn modifies the geometry and gravitational waves are produced when the stars pass through each other.  

Moreover, unlike the case for the black hole collision, where there is a generic form of the gravitational wave, as was shown in \cite{Palenzuela:2006wp} for $\ell=0$ boson stars and in \cite{Jaramillo:2022zwg} for $\ell=0, 1$
boson stars, there is a wide variety of waveform profiles in the case of $\ell$-boson star encounters. These profiles depend on several parameters, being the compactness of the stars one of the most determining. Indeed, for larger values of the compactness the waveform has a profile very similar to the one generated by the black hole collision with the usual phases of plunge and ring-down, whereas for smaller values of the compactness, the waveform generated can be directly identified as not due to a collision of black holes even {\it ad oculum}, presenting a profile with a peculiar bouncing feature, without a sharp peak nor a ring down. 

\begin{figure} [H]
    \centering    \includegraphics[width=\textwidth] {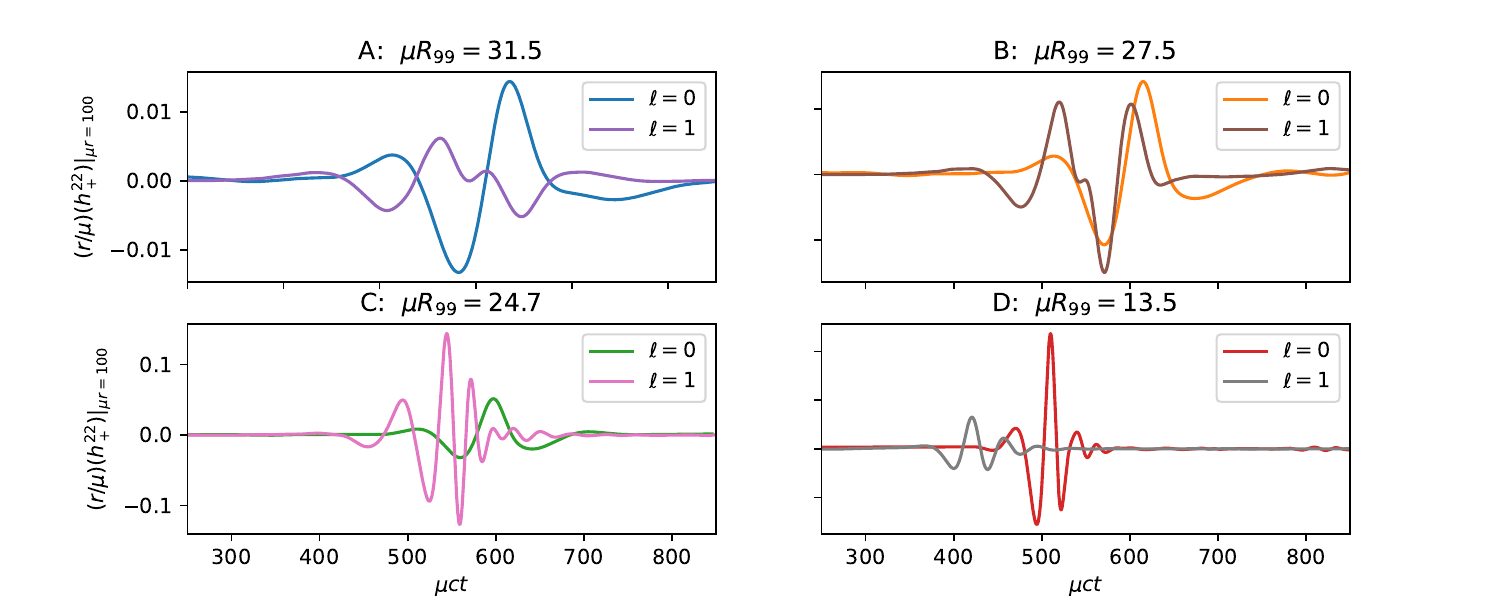}
    \caption{Real part of the 4-scalar Weyl  $Re(\Psi)^{2,2}$, mode $l,\rm{m}=2,2$. The signals correspond to numerical $\ell=0,1$ boson star binaries that head-on collide with different mass configuration.
    }\label{fig:Psi4}
\end{figure}

In Table \ref{tab:models} and Fig.~\ref{fig:Psi4}, we present eight configurations and the corresponding profiles of the real part of the Weyl scalar $\Psi_4$ (see \cite{Degollado:2011gi} for definitions and a review on the null tetrad formulation). In order to compare the results for the $\ell=0$ boson stars with those of the $\ell=1$ boson stars, we analyze cases where the compactness is the same for both set of configurations. Notice that the other parameters, mass $\mu\,M_0$ radius, $\mu\,R_{99}$, and frequency $\omega$ differ for each $\ell$ case. In Fig.~\ref{fig:Psi4} we plot the different waveform profiles for each case and it can be seen how different they are for the different configurations. 

\begin{table}[!ht]
	\centering
	\begin{tabular}{ |c|c|c|c|c|c| } 
	\hline
	Model &  &  &  &  & Remnant  \\
 (waveform identifier) & $\mu \, M_0$  &  $\mu R_{99}$ & $\mathcal{C}$  & $\omega / \mu $ & type   \\
	\hline \hline
        \multicolumn{6}{ |c| }{Boson stars with $ \ell=0$} \\
        \hline
        h$\ell$0A & 0.296 & 31.5 & 0.0221  & 0.985 & BS \\
        h$\ell$0B & 0.333 & 27.5 & 0.0282 &  0.980 & BS \\
        h$\ell$0C & 0.364 & 24.7 & 0.0391 &  0.976 & BS \\
        h$\ell$0D & 0.548 & 13.5 & 0.0838 &  0.931 & BH \\
        \hline 
        \hline
        \multicolumn{6}{ |c| }{Boson stars with $ \ell=1$} \\
        \hline
        h$\ell$1A  & 0.697 & 31.5 & 0.0221 &  0.970 & BS  \\
        h$\ell$1B  & 0.775 & 27.5 & 0.0282 &  0.962 & BS  \\
        h$\ell$1C  & 0.837 & 24.7 & 0.0391 &  0.954 & BH  \\
        h$\ell$1D  & 1.17  & 13.5 & 0.0838  &  0.883 & BH  \\
        \hline 
	\end{tabular}
	\caption{Two kind of boson star models, for $\ell=0$ and $\ell=1$, where $M_0$ is the mass of each initial star, $R_{99}$ corresponds to the radius that contains $99 \%$ of such mass, $\mathcal{C}$ is the compactness, and $\omega$ is the boson frequency. The remnant of the collision can be a boson star or a black hole. All models were obtained in Ref. \cite{Jaramillo:2022zwg}.}
	\label{tab:models}
\end{table}

The models obtained in Ref.~\cite{Jaramillo:2022zwg} made use of numerical simulations performed with the \texttt{Einstein Toolkit} \cite{Loffler:2011ay} and based in previous works handling multiple scalar fields \cite{Cunha:2017wao,Sanchis-Gual:2021edp,Jaramillo:2020rsv}. The Einstein-Klein-Gordon equations are integrated in time using the Baumgarte-Shapiro-Shibata-Nakamura (BSSN) formulation \cite{Shibata95,Baumgarte:1998te}. The geometrical quantities are evolved using the Method of Lines (MOL) of the \texttt{MoL} thorn to solve the equations, using a fourth-order Runge-Kutta scheme provided by the \texttt{McLachlan} thorn ~\cite{brown2009turduckening,reisswig2011gravitational}. 

From the several cases analyzed in \cite{Jaramillo:2022zwg}, we have chosen all the cases in such a way that the stars before the collision are well inside the stable branch, so are stable configurations. The labeled ``A" cases are such that the mass is less than half the maximum mass for the stable configurations, the ``B" cases are equal to half that maximum mass, configurations ``C" are slightly larger than that half and ``D" are close to the maximum value of the mass for stable configurations. The final result after the collision is that cases ``A" and ``B" form a again a boson star, or at least a confined state, for cases ``C" in the $\ell=0$ boson star configurations the final configuration is again a $\ell=0$ boson star; however, in the case for the $\ell=1$ boson stars collision, a black hole is formed after the collision, and in both cases ``D", the collision forms a black hole. 

In Fig.~\ref{fig:Psi4}, we present the waveform (the real part of the $\Psi_4$ Weyl scalar, see bellow), for the four cases for each $\ell$. We plot in the same graphic the corresponding case for each $\ell=0$ and $\ell=1$ boson star collisions, in order to stress the important role of the $\ell$ parameter during the collision. The cases where a black hole is formed $h\ell\,1C,\ h\ell\,0D,\, h\ell\,1D$ are directly seen to have the usual black hole formation profile: the plunge followed by a ring down. The other cases, however, have a different shape that, as mentioned above, even {\it ad oculum} can be seen that do not correspond to black hole collisions.

\subsection{Gravitational Waves calculations} \label{sec:GWcalc} 

In compact numerical binary mergers, it is common to extract gravitational radiation from the mentioned above Newman-Pensore scalar $\Psi_4 \equiv -C_{\mu\nu\sigma\tau} n^\mu \bar{m}^\nu n^\sigma \bar{m}^\tau$, where $C_{\mu\nu\sigma\tau}$ is the Weyl tensor and $l^\mu,\,n^\mu,\,m^\mu,\,\bar{m}^\mu$ define a null vector tetrad \cite{Penrose:1962ij}.
According to the peeling theorem, $\Psi_4$ is the component of the Weyl tensor that falls-off as $1/r$ from the source and corresponds to outgoing gravitational radiation \cite{Bishop:2016lgv}. Radiation $\Psi_4$ is decomposed into multipoles using spherical harmonics (here we use the $l$ instead of $\ell$ to avoid confusion with the background angular momentum parameter) with spin weight $_{-2}Y_{lm}(\theta,\varphi)$:

\begin{equation}
    \Psi_4 (t,r,\theta,\varphi)=\sum_{lm} \Psi_4^{lm}\,_{-2} Y_{lm}(\theta,\varphi).
\end{equation} 
For head-on collision of $\ell$-boson stars, it can be shown, and seeing during the evolution
of head on collisions
that the dominant mode is $\psi_4^{22}$. 

Indeed, the Petrov curvature scalar $\Psi_4$ is related to the  amplitude of the components of the gravitational waves in the TT-gauge: $h_+,\,h_{\times}$ \cite{Teukolsky:1973ha, Maggiore2007}, as
\begin{equation}\label{eq:psi4h}
    \Psi_4 = \frac{\partial^2 (h_{+}-i\,h_{\times}) }{\partial \rm{t}^2},
\end{equation}
an expression valid for each $lm$ mode, and the total expression will be the sum over all the modes:  
\begin{equation}
    h \equiv h_{+}-ih_{\times} = \sum_{l=2}^{\infty} \sum_{m=-l}^l (h_{+}^{lm}-ih_{\times}^{lm})_{-2}Y_{lm}(\theta,\varphi).
\end{equation} 
 Currently, one of the best methods to obtain the amplitude components of gravitational waves is via a {\it fixed-frequency integration} (FFI) \cite{Reisswig:2010di,Reisswig:2010cd}. The terms $\Psi_4^{lm}$ are Fourier transformed considering a step function for small frequencies. In this way, instead of performing a double integral over time, a division is carried out, and through an inverse Fourier transform, the terms $h_+^{lm}(t),\,h_\times^{lm}(t)$ are recovered as time series. 

As long as frontal collisions were studied, the Petrov scalar $\Psi_4$ does not have an imaginary part; it is a real function and thus, from Eq.~\eqref{eq:psi4h}, $h_\times^{lm}(t)=0$.

In general, the response function of the gravitational wave detector to the planar gravitational wave is a linear combination of the two polarizations \citep{Maggiore2007} and is given by
\begin{equation}\label{eq:superposition}
        h(t) = F_+h_+(t) + F_\times h_\times(t)~,
\end{equation}
where $ F_+, F_\times $ are detector's antenna pattern functions  that depend on the sensitivity to each polarization. For an order of magnitude estimate, the gravitational wave strain can be approximated as $h(t) \approx \sqrt{h_+^2(t)+h_\times^2(t)}$. 

In gravitational wave physics, hrss (root sum square) is a measure of the strength of a gravitational wave signal giving by
\begin{equation}
    \text{hrss}=\sqrt{\int (h_+^2(t) + h_\times^2(t)) \, dt},
\end{equation}
this function aids in determining the total strength of the signal observed by detectors such as LIGO and Virgo.

The Fourier transform of the interferometer response to the dimensionless gravitational wave strain become,
    \begin{equation}\label{eq:fourier}
        \tilde{h}(f) = \sqrt{\tilde{h}_+^2(f)+\tilde{h}_\times^2(f)}~.
    \end{equation}

To compute the total energy, $E_{GW}$, carried in a gravitational wave signal, we need to integrate the flux. Assuming an isotropic emission, $E_{GW}$ is related to hrss as \citep{Chatziioannou:2017tdw}
    \begin{equation}\label{eq:E_GW}
        {E}_{\rm GW} = \frac{c^3}{G}\pi^2 d ^2\int_{-\infty}^{\infty}\left(\lvert \tilde{h}_\times(f)\rvert^2+\lvert \tilde{h}_+(f)\rvert^2\right)f^2df~,
    \end{equation}

where $d$ is the distance to the source. 
To characterize the sensitivity for detection at a given frequency, one defines the characteristic strain as \citep{Moore_2014}
    \begin{equation}\label{eq:hc}
        h_c(f) = 2f\lvert \tilde{h}(f)\rvert~.
    \end{equation}
    
The signal to noise ratio, $snr$,  is the inner product of the strain, whitened by power-spectral density of the noise $ S(f) $ \citep{Moore_2014}:
\begin{equation}\label{eq:SNR}
    \text{snr} = \sqrt{\int \frac{4 |\tilde{h}(f)|^2}{S(f)}}df~.
\end{equation}

These expressions will be used in characterizing a gravitational wave comming from the collision of a binary system of $\ell$-boson stars, and will be described in the next section, \ref{Sec:results}

\subsection{Physical parameters and unit system} \label{sec:PhysParUn}

It is a remarkable feature of the boson star configurations that the scalar parameter $\mu$ is a free parameter: the radial coordinate, $r$, the frequency parameter, $\omega$, the mass function, $M$ can be made dimensionless with $\mu$: $r\,\mu, \omega\,c\,\mu, \mu\,\frac{M\,G}{c^2}$ \cite{Jaramillo:2022zwg}. An interesting property of such families of boson star configurations is can be seen in the plot of the total mass, $G\,\mu\,M/c^2$ against the frequency, $\omega/(c\,\mu)$, in the right panel of Fig.~\ref{fig:caracolito}. The snail-like draw describes the family of $0$-boson stars configurations for each frequency, lower curve, and of the $1$-boson stars configurations, upper curve. 

For concreteness, we will focus on the $0$-boson star family, and simmilar description and consequences can be made for other $\ell$-boson star families. The points of the curve located at the right of the maximum describe configurations which are stable under perturbations, whereas those located at the left of the maximum are unstable and under perturbations can migrate to stable ones, dissipate, or collapse unto a black hole (see for example \cite{Liebling:2012fv},  )\cite{Seidel:1993zk}).
Another important parameter of boson stars is the defined compactness ${\cal C}$, the ration of the mass of the star to its size, Eq.~\eqref{eq:compacidad}, at a given radius; it is usually considered the ratio between the mass $M_{99}$ to the size $R_{99}$:
\begin{equation}
{\cal C}=\frac{G\,M}{R\,c^2}, \label{eq:compacidad}   \end{equation}
and is a untiless parameter, the subindex $99$ is usually droped, and its is presented for the family of configurations for each frequency in the inset of the right panel of Fig.~\ref{fig:caracolito}. 

The plots of the total mass and the compactness are very useful to recover the physical mass and size of a configuration for a given value of the mass of the scalar field, $m_\phi$. We describe the procedure as follows.

To determine a physical case, choose a given configuration which means select a value $\omega_0/(\mu\,c)$, and read the point, $P_0$, in the plot of mass M versus frequency, Fig.~\ref{fig:caracolito}, left panel, say for the $\ell=0$ case, so that: 
 \begin{equation}
 \frac{G\,\mu\,M}{c^2}=P_0,    
 \end{equation}
for a given frequency, say $\frac{\omega_0}{c\,\mu}$
so that $P_0$ is a value we can read from the graph, values for instance from the second column in Table \ref{tab:models}, usually or order $\sim 0.2, 0.5$.
Adding the corresponding constants, $\mu=\frac{m_\phi\,c}{\hbar}$, we see that this is a ratio of lengths, and $P_0$ is a dimensionless quantity:
 \begin{equation}
 \frac{m_\phi\,c}{\hbar}\,\frac{M\,G}{c^{2}}=\frac{m_\phi\,c^{2}}{\hbar\,c}\,\frac{M\,G}{c^{2}}=P_0,  \label{eq:P0}  
 \end{equation}
where we multiply and divided by $c$ in order to give the value of $m_\phi\,c^{2}=E_\phi$, usually given in eV, and $m_\phi$ is the rest mass of the boson star. Thus, we obtain for the total mass
 \begin{equation}
 M=\frac{P_0}{m_\phi\,c^{2}}\,\frac{\hbar\,c^{3}}{G}, 
 \end{equation}
whith $G = 6.67384\times 10^{-11}\,\mathrm{m}^3\mathrm{kg}^{-1}\mathrm{s}^{-2} =1.34\times 10^{11} M_{\odot}^{-1} \mathrm{km}^3 \mathrm{s}^{-2} $,  $ c = 2.99\times10^{8}\,\mathrm{m\;s}^{-1} =  2.99\times10^{5}\,\mathrm{km\;s}^{-1}$ and  
$\hbar=6.58\,\times\,10^{-16}\,\rm{eV\,s}$, 
with $M_\odot$ stands for solar masses. Thus, the total mass M become
 \begin{equation}
 M=1.3126\,\times\,10^{-10}\,\frac{P_0}{m_\phi\,c^{2}\,[\rm{eV}]}\,\,\rm{eV} M_\odot, 
 \end{equation}
and considering that $m_\phi\,c^{2}=10^{-n}\,\rm{eV}$, where tipical values for $n$ are $n=-9$
for the Higgs boson, $n=5$ for the axion, $n=22$ for the ultralight scalar field, also known as fuzzy dark matter, we obtain for the mass of the boson star under consideration that
 \begin{equation}
 M=1.3126\,\times\,10^{n-10}\,P_0\,M_\odot.
 \label{totalmass}
 \end{equation}

Regarding the size of the object, given by $R_{99}$, we can use the plots for the compactness, we use the inset of the right panel of Fig.~\ref{fig:caracolito} . For the chosen frequency, $\omega_0/(\mu\,c)$, we read the corresponding value of the compactness, say $\cal{C}_0$, for instance values in the third column in Table \ref{tab:models},
as $\mathcal{C}=\frac{M}{R}$, we have, with the corresponding units in order to have a dimensionless compactness:
\begin{equation}
R=\frac{M\,G}{c^{2}\,\mathcal{C}}, 
\end{equation}
which implies that
\begin{equation}
R=\frac{1.4998}{\mathcal{C}}\,\frac{M}{M_\odot}\,\rm{km}, 
\end{equation}
for the mass obtained above in Eq. \eqref{totalmass}:
\begin{equation}
R=1.9674\,\times\,10^{n-10}\frac{P_0}{\mathcal{C}_0}\,\rm{km}. \label{R0bs}
\end{equation}

Considering a $0$ boson star and value for the frequency of $\omega_0/(\mu\,c)=0.985$, which corresponds to $P_0\sim 0.296$. Recall that the $\ell$ boson star configurations are numerical, so that for the exact corresponding values, one construct tables.  The compactness value for the chosen frequency is of $\mathcal{C}\sim 0.0221$. As mentioned in the above discussion, all the configurations scale with the scalar field mass. To complete the determination of a physical case, it is needed to choose a particular scalar field, that is a given mass, $m_\phi$, which is equivalent to choose a value for $n$ for the given configuration, we determine the corresponding total mass of the star and its size. 

In a similar manner, we may choose the total mass of a given $\ell=0$ boson star; say a mass of the order of solar mass and, form Eq.~\eqref{totalmass}, for the chosen $P_0$, we need a mass of a scalar field of $n\sim 10$ that is $10^{-11}\,\rm{eV}$, implies $3.88\,M_\odot$, and, as all the parameters are now fixed, Eq.~(\ref{R0bs}) implies  $R_{99}=263.5\,\rm{km}$. Thus, a solar mass $\ell=0$ boson star made out of a scalar field of $10^{-11}\,\rm{eV}$ has a radius 26 times larger than neutron star with the similar mass. The same total mass but for a lighter/(heavier) scalar field gives a larger/(smaller) boson star.

One may wonder on the size of these objects, could a galactic halo be seen as a gigantic boson star? Let us give a couple of concrete examples. First, notice that there are some constraints that the parameters have to fulfill. As the maximum value that $\mu\,M$ can attain is 0.65, from Eq.~(\ref{eq:P0}), we see that
\begin{equation}
1.16\,10^{10+m-n}<1,    
\end{equation}
were we are considering a mass of the boson star given by $10^m\,M_\odot$, and the usual $10^{-n}\,{\rm eV}$  
for the mass of the scalar field. On the other hand, from the graph of the compactness, Fig.~\ref{fig:caracolito}, we see that for a $\ell=0$ boson star the maximum compactness is of the order of $0.11$ thus, from Eq.~(\ref{eq:compacidad}), so that, we obtain a constraint for the size of the $\ell=0$ boson star (bs) given a mass:
\begin{equation}
R_{0-bs}>13.54\,{\rm km}\,10^{m}.    \label{Rmaxbs}
\end{equation}

In this way, we see that if we want to describe a $\ell=0$ boson star with a mass equal to a solar mass, $m=0$, we see that the scalar field must be lighter that $10^{-10}\,{\rm eV}$, and must have a size larger than $13.54\,{\rm km}$; if we would like it to have the solar radius, $R_\odot=6\,\times\,10^5\,{\rm km}$, then the compactness must be ${\cal C}_\odot=2.48\,\times\,10^{-6}$, and if we are dealing with a scalar field with mass $10^{-11}\,{\rm eV}$, then, $P_\odot=0.076$. On the other hand, if we want the $\ell=0$ boson star to be halo like, then $M_h=10^{12}\,M_\odot$, and $R_h=10^2\,{\rm kpc}=3.08\,\times\,10^{18}\,{\rm km}$, then the compactness of the object is ${\cal C}_h=4.83\,\times\,10^{-7}$, the mass of the scalar field must be lighter than $10^{-22}\,{\rm eV}$, so that for this value, we get again $P_h=0.076$. The final value for the mass of the scalar field is fixed as long as the value for $\omega$ has to be the same for the mass and the compactness, whereas we present here an order of magnitude analysis.

We see that there is a large of freedom on the physical parameters of a $\ell=0$ boson star. For the $\ell=1$ boson star, similar constraints hold, changing only the values for the maximal mass and compactness. We finish this section noticing that when the boson field known from particle physics is considered to form boson stars, as such boson are heavy, with masses or the order of ${\rm GeV}$, in our notation, $n \sim -9$, from Eq.~(\ref{totalmass}), with $P_0=0.63$ for the maximal mass case, we see that the masses obtained using these fields are $8.26\,\times\,10^{-20}\,M_\odot$, asteroids like and were called {\it mini-boson stars}.

As mentioned above, from the numerical simulation we obtain the strain $h$, the data strain as a unitless profile, it is the one emitted during the plunge. As it is well known from the Peeling theorem \cite{Wald:1984rg}, we know that the Petrov scalar $\Psi_4$ is the one that decays less rapidly compared to the other ones, decaying as $1/r$, and in term of the scale that we are using such decay is $1/(\mu\,r)$, thus the magnitude of the curvature scalar, and thus of the strain, at the detection point (label ``$d$"), and consider that the source, label ``$s$" is at a physical distance $d$, we obtain that the detected strain is given by:
\begin{equation}
  h_d =  \frac{1}{\mu d}\,h,  
\end{equation}
and in terms of the mass of the scalar field, given as $10^{-n}\,{\rm eV}$, and considering that the collision occurred at a a hundred megaparsecs, we obtain 
\begin{equation}
    h_d= \frac{\hbar c}{10^{-n}\,{\rm eV}\,d} h=\frac{6.387\times 10^{n-32}}{{d}_{10^2\,Mpc}}\,h. 
    \label{fig:h_d}
\end{equation}
In this last expression, we see what we have mentioned above: the scalar field theory allows the existence of boson stars and their gravitational waves; however, the theory does not constrain the mass of the scalar field, equivalently, nor the $\mu$ parameter. The sensitivity of the gravitational observatory is the one that determines the range of values of the mass of the scalar field, $m_\phi\,c^2$, (equivalently of the parameter $\mu=\frac{m_\phi\,c^{2}}{\hbar\,c}=5.082\,\times\,10^{9-n}\,{\rm{km}}^{-1}$) so that the amplitudes and frequencies of gravitational waves fall within the observation ranges of the gravitational detectors. From the expression 
 \eqref{fig:h_d}, we see that, as the detector sensitivity for the strain is about $10^{-22}\,$, and this impose a constraint on the mass of the scalar field composing such binary system. 

Indeed, with the actual magnitude of the perturbation at the emission $h$, it can thus be determined the mass of the scalar field which formed the masses of the $\ell$-boson stars, $\ell=0,1$, colliding at a hundred Megaparsec from us, which can be detected by LIGO.

This analysis sets a limit for the lightness that the mass scalar field requires in order that the gravitational waves generated by it are detected. Heavier scalar fields form less massive stars, see Eq.~(\ref{totalmass}), so that the strain of the gravitational wave generated under these conditions might not be enough to be detected by LIGO. Finally, notice also that the strain analysis does not have a lower limit for the mass of the scalar field. 
\begin{figure} [H]
    \centering    \includegraphics[width=\textwidth] {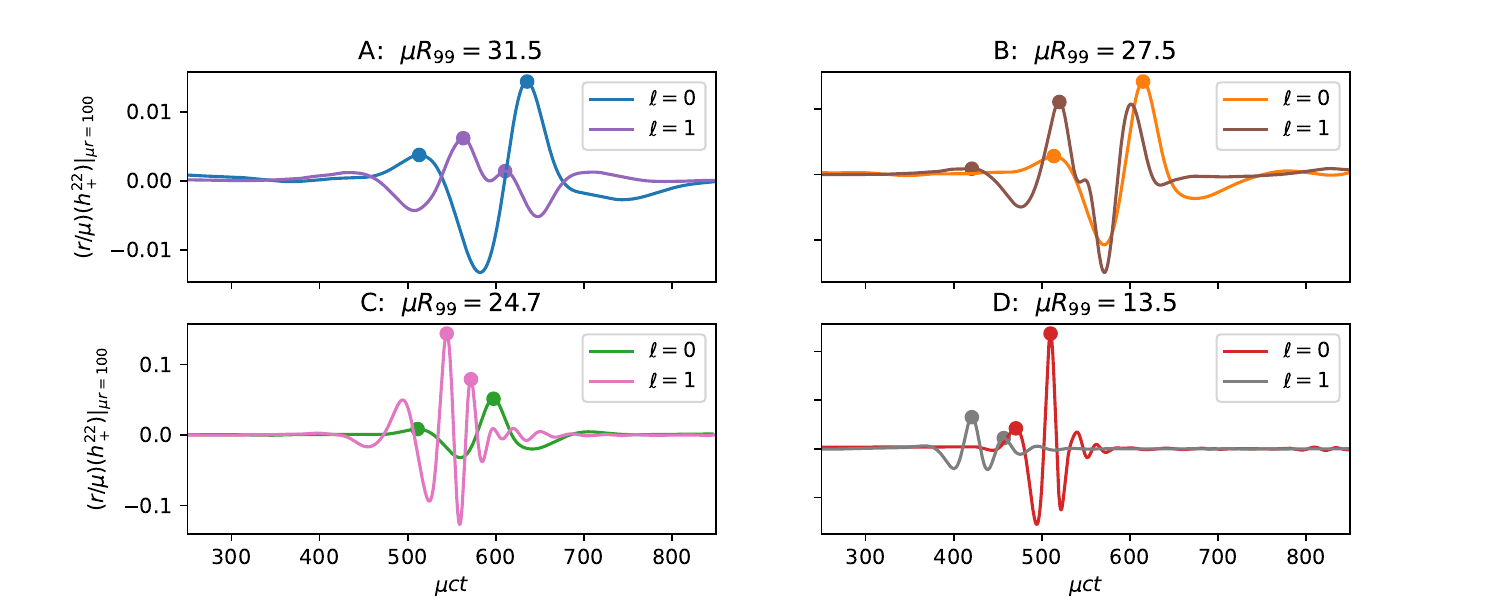}
    \caption{Gravitational wave amplitude for the models $h\ell0A$ to $h\ell0D$ and $h\ell1A$ to h$\ell$1D in dimensionless  units. Where we have chosen values for the product $\mu\,R_{99}$ and fixed the detection distance, so that the strain is already scaled. Also, we have marked the nodes in the waves which help us to compute the wave lenghts.}\label{fig:h0_fp}
\end{figure}

In the following section, we present concrete examples of wave profiles for several head-on configurations analyzed in \cite{Jaramillo:2022zwg}, and determine the maximum value for the mass of the scalar field that generates detectable gravitational waves. We also present the frequency analysis which allows us to further constrain such scalar field masses, even from bellow.  

\begin{table}[!ht]
	\centering
	\begin{tabular}{|c|c|c|c|c|c|} 
	\hline
	   Model & &  & &   &\\
    (waveform identifier) & $ \hat{h}_0$  &  $\hat{\lambda}$  & $1/\hat{\lambda}=\hat{f}_{\mathrm{peak}}$ & $n_{min}$ & $n_{max}$ \\
	\hline \hline
           \multicolumn{6}{ |c| }{Boson stars with $ \ell=0$}  \\
        \hline
           h$\ell$0A & 0.014 & 122.5 & 0.008 &12.88& 10.88\\
           h$\ell$0B & 0.0284 & 101.25 & 0.009 &12.96&10.96\\
           h$\ell$0C & 0.051 & 86.25 & 0.011 &13.03&11.02\\
           h$\ell$0D & 0.236 & 39.37  & 0.025 &13.29&11.29\\
        \hline 
        \hline
           \multicolumn{6}{ |c| }{Boson stars with $ \ell=1$}  \\
        \hline
           h$\ell$1A  & 0.006 & 47.5 & 0.021 &13.29&11.29\\
           h$\ell$1B  & 0.022 &   99.37 & 0.010 &12.97&10.97\\
           h$\ell$1C  & 0.144 &  27.5  & 0.036 &15.53&11.53\\
           h$\ell$1D  & 0.065 &  36.25  & 0.027 &13.41&11.41\\
        \hline 
	\end{tabular}
	\caption{Dimensionless values, denoted by hat $\hat{}$, for the maximal amplitude, and peak frequency for the eight gravitational signals  $h\ell0A-D$ y $h\ell0A-D$. In the final columns, is presented the range of exponents in the scalar field mass, $10^{-n}\,{\rm eV}$ which fits with this values.}
	\label{tab:tab2}
\end{table}

The fact that the gravitational detectors work in a frequency range between $10$ Hz to $1000$ Hz gives us another constraint on the mass of the scalar field mass that can be detected by means of the passing of the gravitational wave. Indeed, from the waveforms presented in Fig.~\ref{fig:h0_fp},  we see that the wave length $\hat{\lambda}$ of the signals varies approximately from $125$ the longest to $35$ the shortest; the actual values are given in Table~(\ref{tab:tab2}). As the scale length is $\mu$, which actually is the inverse of the characteristic wavelength of the scalar field, we have for such wave-lengths (recall that they are normalized by the $\mu$ parameter, so we give back the units)
\begin{equation}
\lambda=\frac{\hat{\lambda}}{\mu}=\frac{m_\phi\,c^2\,\hat{\lambda}}{\hbar\,c}=\frac{10^{-n}\,{\rm eV}\hat{\lambda}}{\hbar\,c}, 
\end{equation}
so that, as the gravitational waves travel at the speed of light, for the angular frequency we obtain:
\begin{equation}
\omega=\frac{2\,\pi\,10^{-n}\,{\rm eV}}{\hbar\,\hat{\lambda}}. \label{omega_gw}   
\end{equation}

As discussed in more detail in the next section, the LIGO frequency range for the gravitational waves detection is between ten and a thousand Hertz, a fact that, with the expression above, allows one to determine the mass of the scalar field composing head one collision of the $\ell$-boson stars generating gravitational waves that can be detected by LIGO. Indeed, from Eq.~(\ref{omega_gw}) we see that:
\begin{equation}
n=\log\left(\frac{2\,\pi}{\hbar\,\omega\,\hat{\lambda}}\right) = 15.97 - \log\left(\omega_{\rm LIGO}\right) -
\log\left(\hat{\lambda}\right).     
\end{equation}
Recall that $n$ labels the mass of the scalar field, $m_\phi\,c^2=10^{-n}\,{\rm eV}$. In the actual profiles presented in Fig.~\ref{fig:h0_fp}, we also have the amplitude of the strain at the emission, $h$ (actually, to be precise, the gravitational waves are measured at $\hat{r}=100$). 

With this information, we are thus able to determine the range of the mass of scalar field generating detectable gravitational waves.  A word of caution has to be given: even at the outset of the configurations, there is still some degeneracy on the values assigned to a given configuration and, extra parameters as
in the case of $\ell=1$ boson star, there is an extra freedom in the orientation of the assembling of one of the stars with respect to the other. As discussed in \cite{Jaramillo:2022zwg}, the waveforms depend on these parameters as well. So that we present here possible cases and a question of principle that it is possible to generate and detect gravitational waves coming from collisions of $\ell$-boson stars and stress the main features of such task. 

For the present work, we have chosen data sets of $\ell$-boson stars corresponding to the initial configurations of two stars of the same type, positioned along the stable branch. The simulation represents aligned stars; both stars have the same orientation. The cases shown in our analysis correspond to two spatially separated lumps formed from the same fields, often referred to as coherent states.

For the $\ell= 0$ boson stars merger, the difference between the diagonal components averages to zero, confirming the tendency to sphericity. For the case $\ell=1$, the tendency is non-sphericity. At the end state in the merger of the $\ell= 0$ boson stars it tends towards a new $\ell= 0$ boson star, whereas for the case $\ell= 1$ merger it does not happen, though it continues to be a bound state of the scalar field. It is possible that the asymptotic end state is a localized configuration with fewer symmetries; the simulations performed in \cite{Jaramillo:2022zwg}, however, can only raise this possibility, not establish it.

\section{Results}     
\label{Sec:results}


As discussed in the previous section, the gravitational signals generated by the plunge of boson stars have a wide spectrum of shapes, ranging from those very similar to the ones produced by the collision of black holes to those clearly different. In this section, we describe in detail the constraints set by the LIGO's present day capabilities on the mass of the scalar field conforming the colliding stars whose gravitational waves could be detected. We will consider boson star configurations similar in mass and distance to the ones that have been detected for black holes, in order to make the study easier, by comparing with a well understood signal. Our aim is to give a general idea on how some of these different profiles look in the detector, with the ones similar to the black hole collision showing well known shapes, and describing the features in the detection of the gravitational waves clearly different from them. 

\subsection{Analysis of all signals in a scalar field mass range}
\label{sec: All signals various mu}

The data processed in this section are obtained from Ref. \cite{Jaramillo:2022zwg}; the graphs use the values of the physical quantities of the speed of light and the universal gravitational constant, as given in Section \ref{sec:PhysParUn}.  
Moreover, as discussed at the end of Section \ref{sec:GWcalc}, 
the parameter allows us to detect waves coming from the collision of scalar field with masses in the interval $m_\phi c^2\in (5.9\times10^{-15},1.8\times 10^{-10})\, {\rm eV}$, that is for the parameter $\mu=m_\phi c^2/\hbar$ in the range: $3\times10^{-8}\,{\rm m}^{-1}< \mu <9\times10^{-4}\,\rm{m}^{-1}$. As mentioned above, we consider the distance to the source is $d=100$ Mpc in order to compare with similar wave profiles coming from black holes at such a distance.

Following our aim, {\it i. e.} to analyze the gravitational waveforms coming from the binary boson system, we explain in subsection \ref{sec:GWcalc}, the strain characteristic, and the signal noise ratio that allow us to determine if the signals are present in the detection LIGO range in the future O5 run. 
In particular, we use the eight boson stars models (h$\ell$0A, h$\ell$0B, h$\ell$0C, h$\ell$0D, h$\ell$1A, h$\ell$1B, h$\ell$1C, h$\ell$1D) explained in Fig. \ref{fig:h0_fp} and Table \ref{tab:models}. 
 
In this section we describe how to choose the parameters so that the wave forms generated by the $\ell$ boson stars plunges, be detected by the gravitational observatories, considering that the source system is located at a distance of 100 Mpc. As a point of comparison, these GW characteristics are also shown for the case of a binary black hole with equal masses of $10\,M_{\odot}$ and at the same distance as the boson star (see the dotted horizontal black lines in all plots of this section).

\begin{figure}[H]
    \centering
    \begin{tabular}{cc}
    \includegraphics[width=0.49 \textwidth]{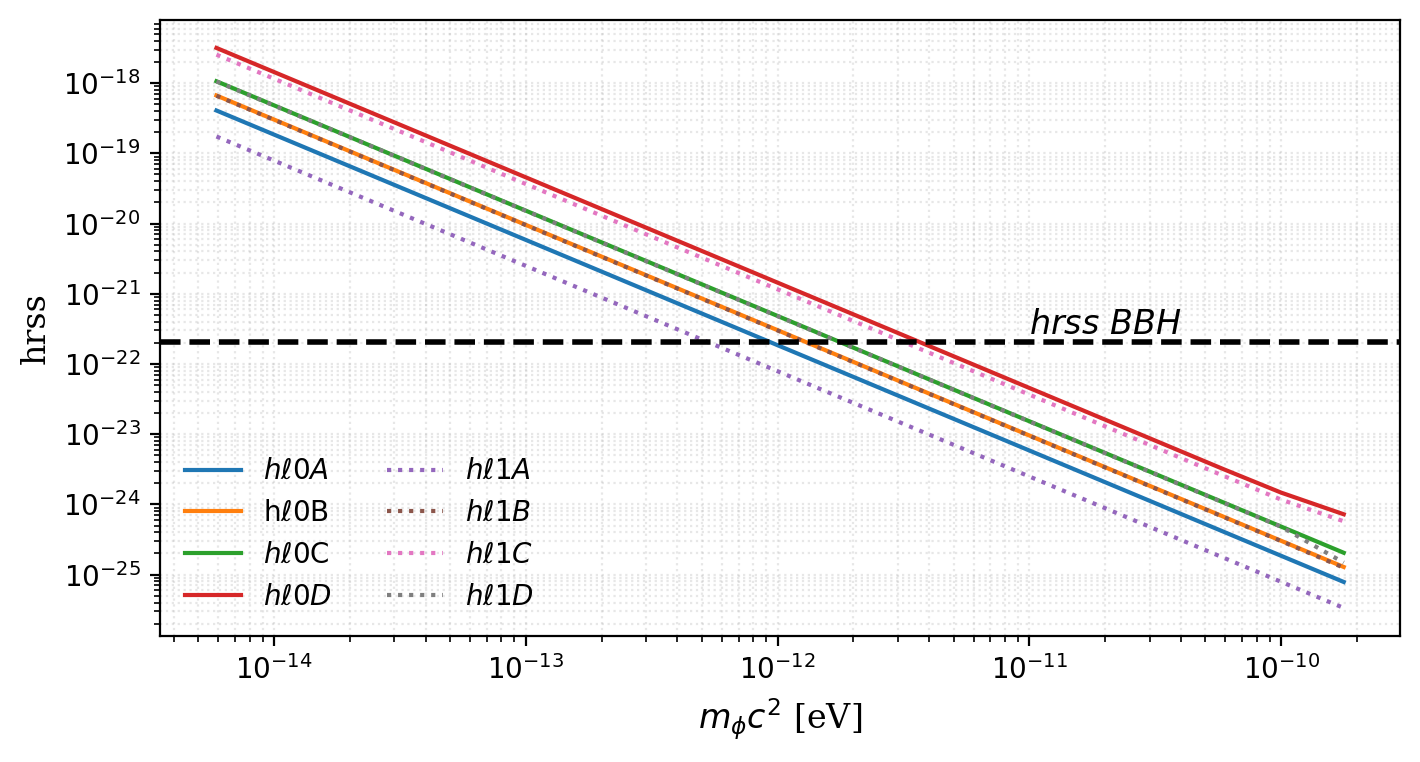}
    &
    \includegraphics[width=0.49 \textwidth]{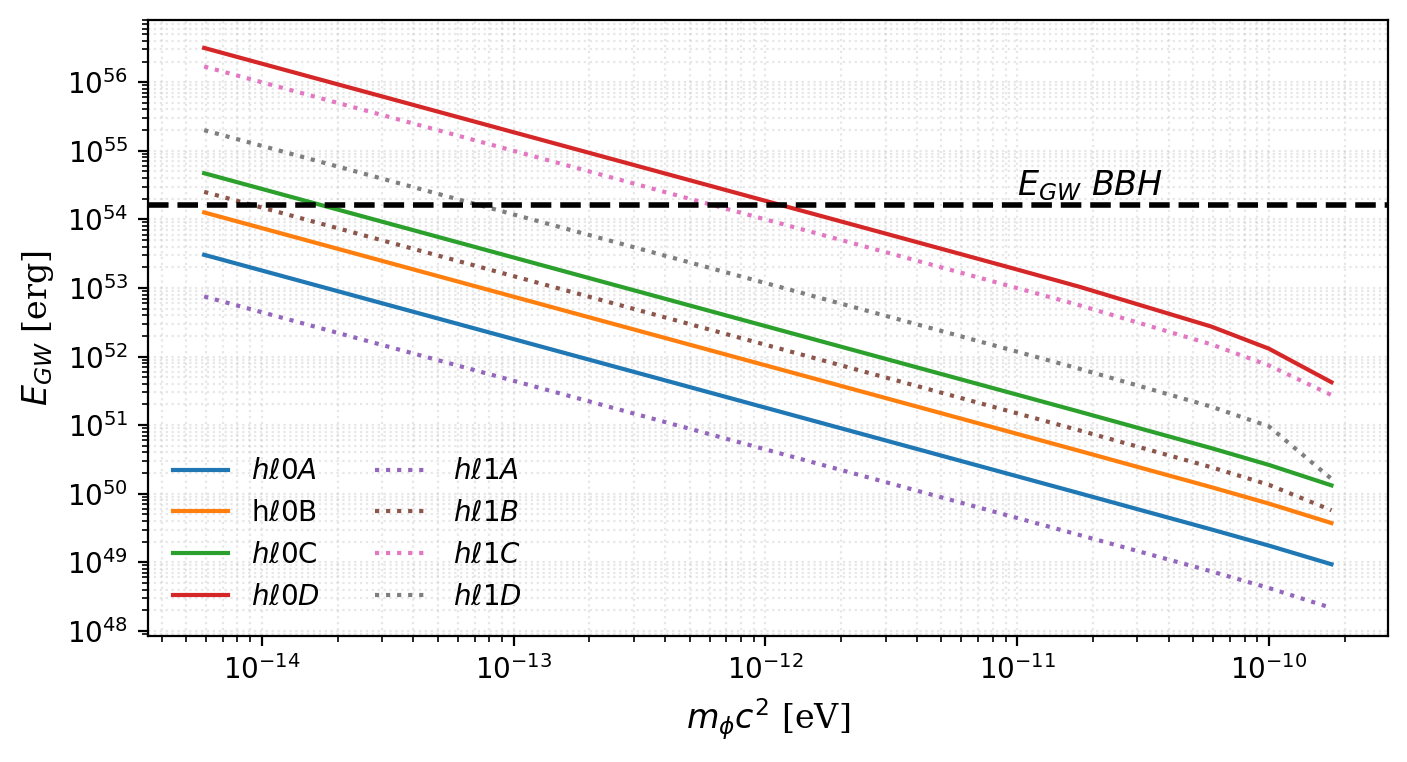}
    \\
    (a) & (b) \\
    \includegraphics[width=0.49 \textwidth]{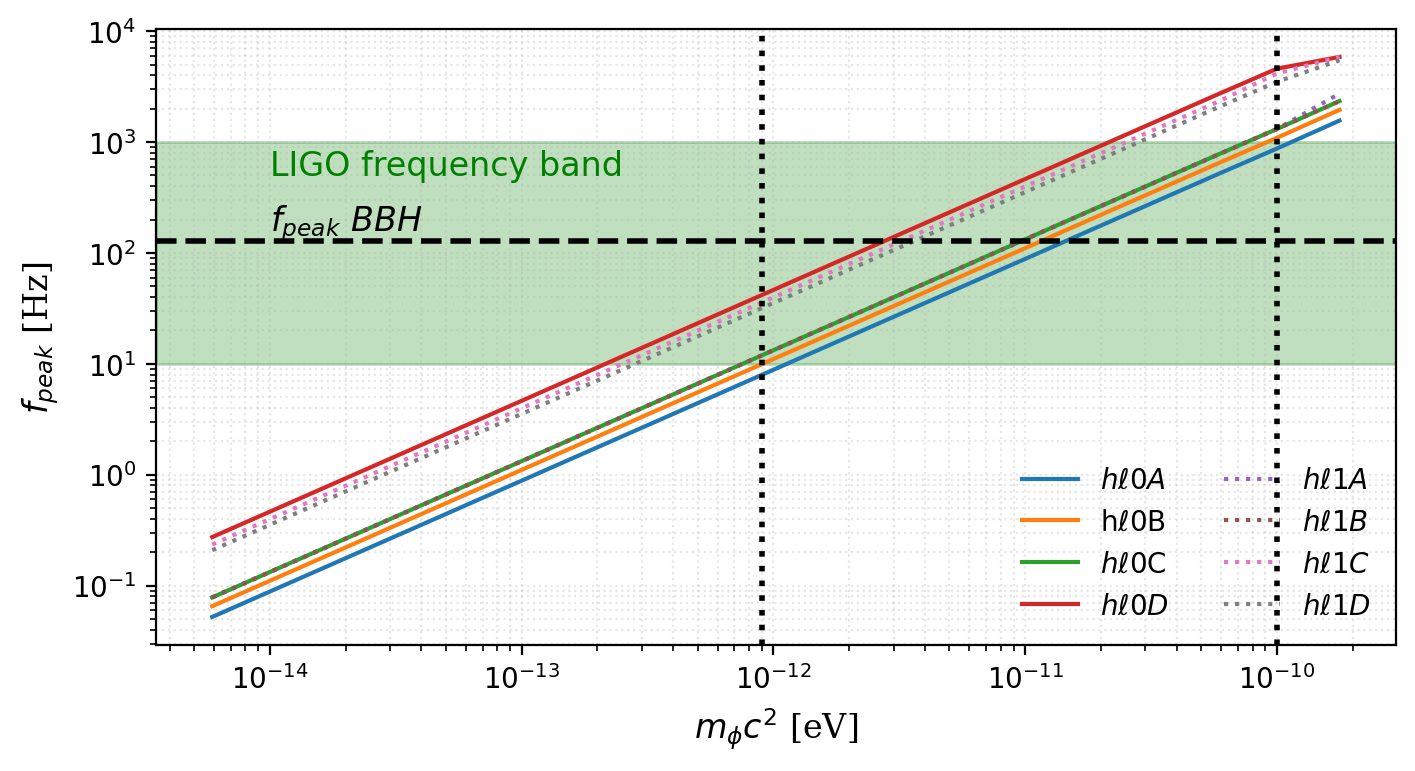}
    &
    \includegraphics[width=0.49 \textwidth]{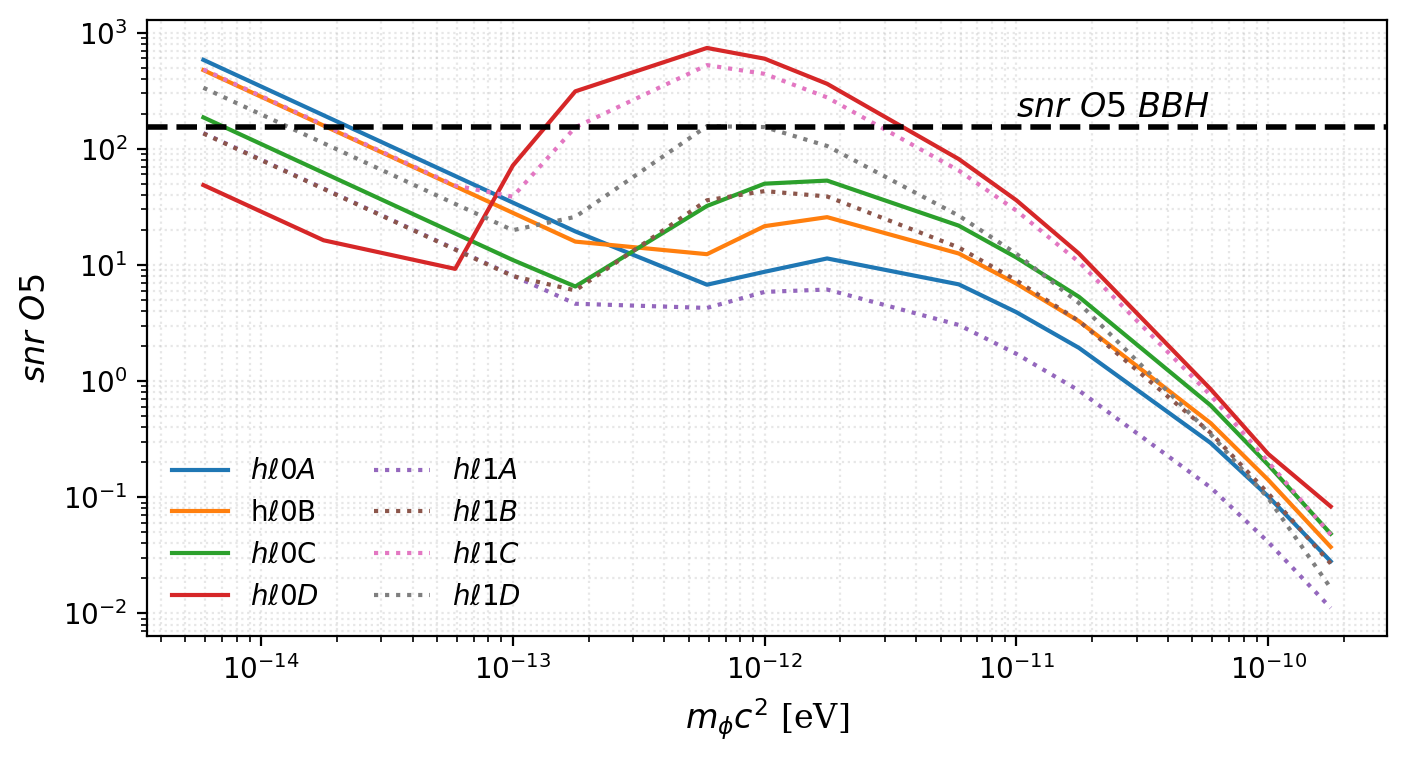}
    \\
    (c) & (d) 
    \\
    \end{tabular}
    \caption{Variation of the $\ell$-boson stars GW signal characteristics (a) hrss, (b) energy, (c) peak frequency and (d) $snr$, as a function of the mass of the scalar field $m_\phi c^2$ in units of electron volts. 
    For reference, these gravitational waves characteristics are also shown for the case of a binary black hole with $m_1 = m_2 = 10\,M_{\odot}$ at a distance of $d$=100 Mpc (see dashed horizontal black lines in all plots).
    For the case of the peak frequency, the shaded green patch indicates the frequency band of the LIGO detector, and we have included vertical lines to mark the scalar field masses which produced events that could be {\it hear} by the LIGO detector.
    } 
\label{fig:several_mu}
\end{figure}

In Fig. \ref{fig:several_mu} (a), we relate the root-sum-square amplitude hrss of the gravitational wave with the mass-energy $m_\phi\, c^2$ of the scalar field composing the stars, and we have included a dashed line marking the corresponding value of a binary black hole (BBH) of equivalent masses, which is around $10^{-22}\,{\rm m}$, 
for such strain. Notice that events produced by stars with smaller (larger) scalar field masses produce waves with a larger (smaller) strain than the 
one produced by the binary black hole system, coinciding in strain for the case, h$\ell$0A, with scalar field mass of $10^{-12}\,{\rm eV} 
$. 

In Fig. \ref{fig:several_mu} (b) we present the energy of the gravitational waves of the eight events analyzed, $E_{GW}$. Again, we include with a dashed line the corresponding one of a binary black hole of $10\,M_\odot$, which has an approximate value of $10^{47}$ erg.  We see that the energy of the gravitational waves decreases  with the mass of the scalar field, and the case which produces the wave with a similar energy than the ones produced by the balck hole is $h\ell0D$.

In Fig. \ref{fig:several_mu}(c), we plot the frequency peak $f_{peak}$ defined as the frequency at which the gravitational wave spectral energy density $dE_{GW}/df$ reaches its maximum \cite{PhysRevD.93.042002}; the frequency range for binary black hole is around $200 {\rm Hz}$, marked by a dashed line. In this plot, we see that the range of gravitational waves that could be detected by LIGO, 
whose frequency band is in the range of $10 <f_{peak}<10^3$, are those produced by the collision of $\ell$-boson stars formed by scalar field masses in the range of $10^{-12}$ to $10^{-10}$ ${\rm eV}$. Signals produced by $\ell$-boson stars outside this range of masses, can not be detected by present day LIGO architecture, we have included dotted vertical lines to mark such range.

Finally, in Fig. \ref{fig:several_mu}(d) we calculated signal to noise ratio $snr$ of the gravitational waves produced by the $\ell$-boson star, this measures the strength of the stochastic background relative to the random noise in the detectors: a very low $snr$ means that the signal is still buried beneath our detector noise, while a very high $snr$ means we can be very confident that we have made a detection. We plot the $snr$ using O5 data and, again, we include the corresponding value of the black hole system, around $200$. We observe that the $snr$ of the gravitational waves generated by $\ell$-boson stars decreases for smaller scalar field masses, however, they present a bum, an increase in the ratio as the mass of the the scalar field increases, and then decreases again. In particular, there is a maximum for signals produced by scalar field mases of about $10^{-12}\,{\rm eV}$, reaching $snr$ values even larger that the corresponding to the black hole case.
In the four plots Fig. \ref{fig:several_mu}, the model h$\ell$0D has the highest values of strain and energy, and is the one that reaches the largest $snr$. 
On the other hand, $h\ell1A$ has the corresponding lowest values. Also, we are able to see that the h$\ell$0D with scalar mass of $10^{-12}\,{\rm eV}$, is the one that has more similar features to the one of the black hole, while $h\ell1A$ is the case that differs the most from the black hole example. 
In later sections, we will make a comparison between these two signals.

\subsection{Analysis of all signals for a scalar field mass \texorpdfstring{$m_\phi c^2 = 1.0\times10^{-11} \, {\rm eV}$}{1e-11 eV}}


In order to compare the different signals generated by the collisions of the eight h$\ell$ signals presented, we assume in this section that all configurations arise from the same scalar field, defined by the mass $m_\phi c^2$, allowing us to examine and compare the corresponding signals.

\begin{figure}[H]
    \centering
    \includegraphics[scale=0.8]{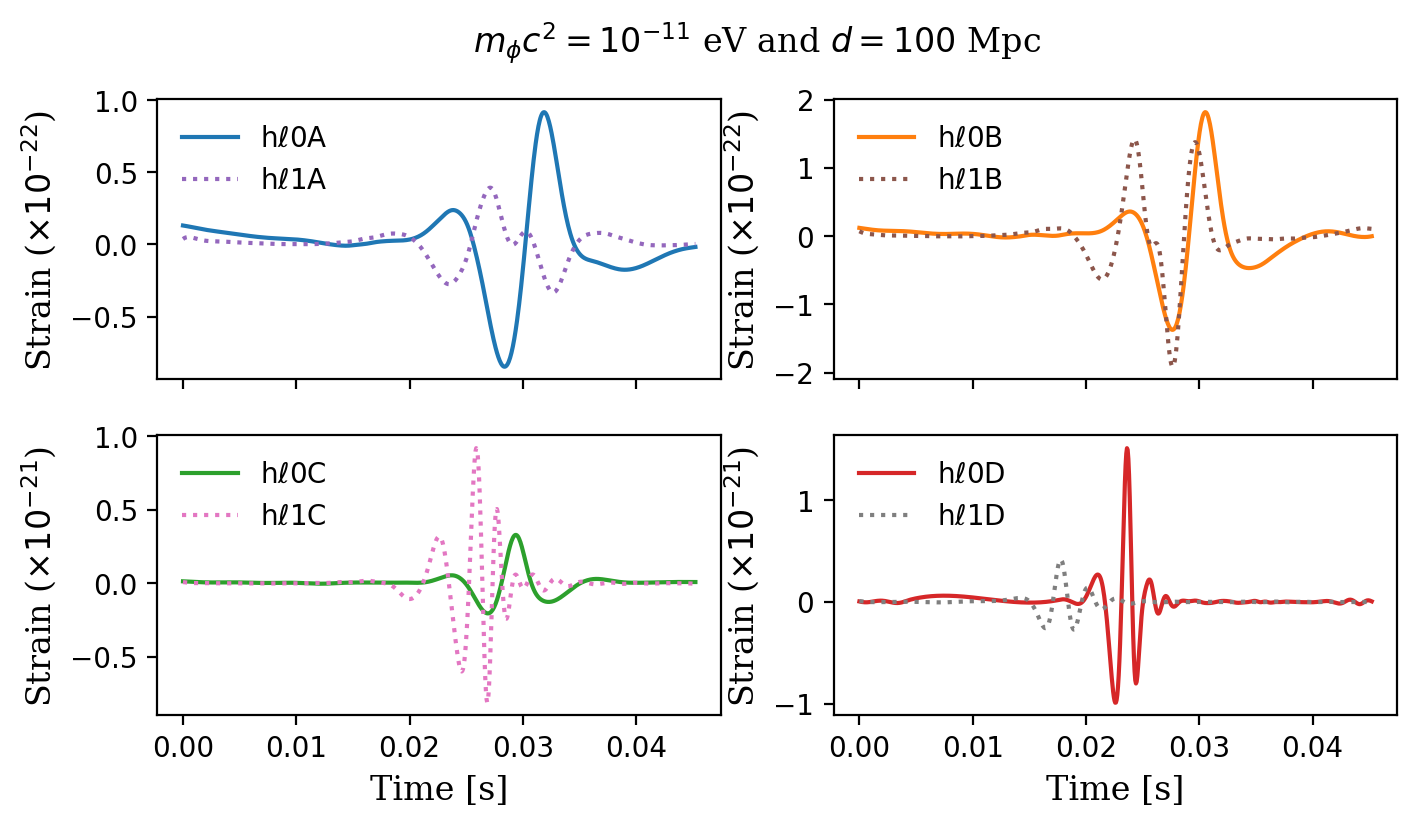}
    \caption{
    Strain of the $l$-boson star GW signals for the case of a scalar field mass of $m_\phi c^2 = 10^{-11}$ eV and a distance of $d$=100 Mpc. $h\ell1C$ and $h\ell0D$ show more similarity with a black hole shape, in their periodicity and amplitude.
    }
    \label{fig:strain}
\end{figure}

 In Fig. \ref{fig:strain} we plot the strain characteristic for the different models of the signals represented by the cases h$\ell$0 and h$\ell$1 for examples A to D; we select the scalar field mass to be around $m_\phi c^2=10^{-11}\,{\rm eV}$ and the distance $d$=100 Mpc according to the analysis in Subsection \ref{sec: All signals various mu}. As mentioned above, the boson star binary system experiences plunge, but no collision occurs, the system, however, emits gravitational waves. The models h$\ell$0A - D have the following initial masses: $6.68,\,7.51,\,8.21$ and $12.36\,M_{\odot}$, respectively; the models h$\ell$1A - D have masses: $15.72,\,17.48,\,18.88\,M_\odot$ and $25.49\,M_{\odot}$, respectively. Fig. \ref{fig:strain} illustrates the eight signals, each demonstrating the following behavior: for $l=0$, we can observe in all of them fewer oscillations than in case $\ell=1$. We can observe that the shape of the waveforms of models h$\ell$0A to h$\ell$0C are similar with different scale. On the other hand, the model which produces a most similar signal to the one of a binary black hole system, is the one represented by $h\ell1C$; also notice that more oscillations are presented during the process. Binary black holes, such as GW150914 \cite{LIGOScientific:2016emj}, radiated around $3\,M_\odot c^2$ in energy during the merge. In general, the energy emitted by the $\ell$-boson star plunge is less than that. %

\begin{figure}[H]
    \centering
    \includegraphics[scale=0.5]{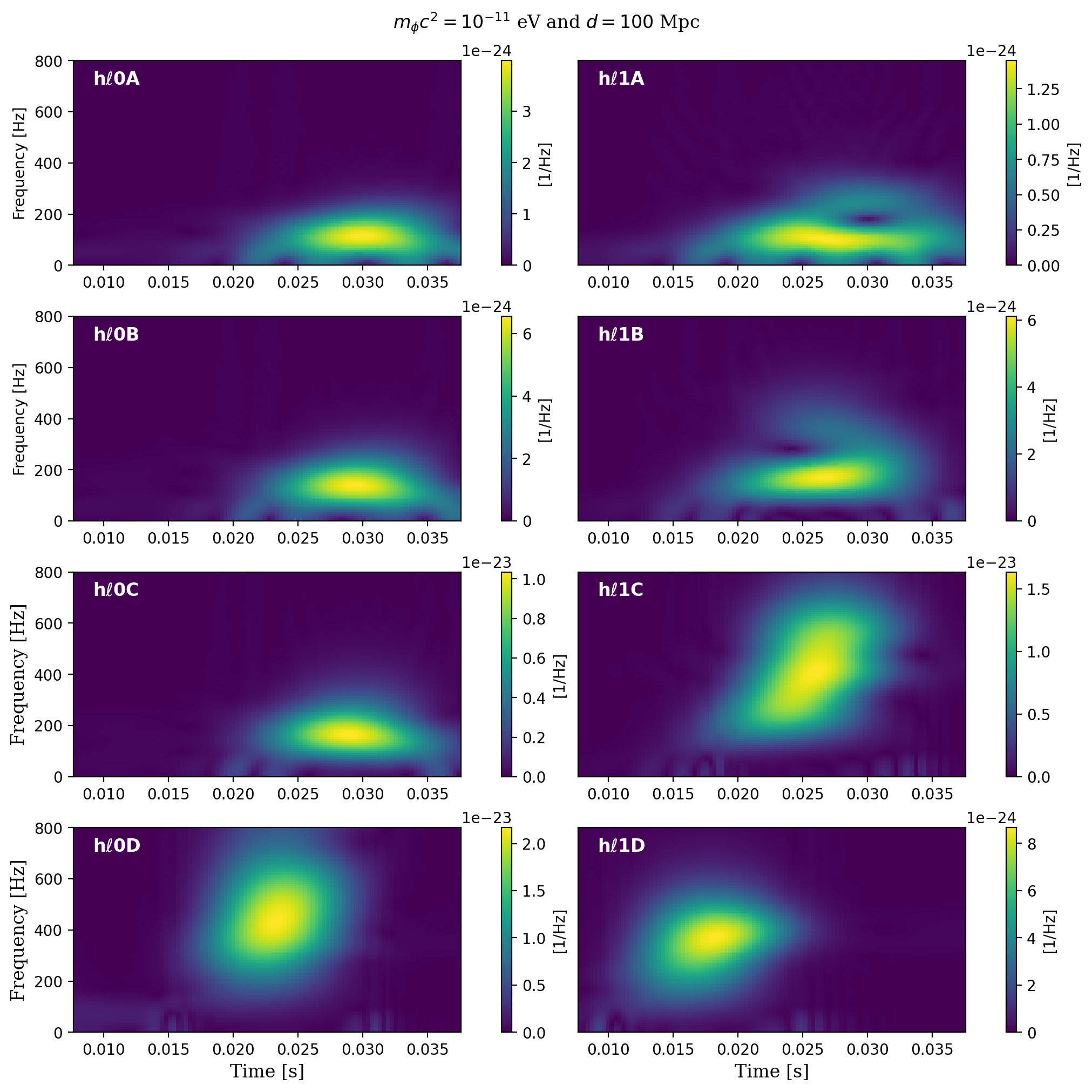}
    \caption{
    Spectrogram of the $\ell$-boson star GW signals for the case of a scalar field mass of $m_\phi c^2 = 10^{-11}$ eV and a distance of 100 Mpc. Many of the plots show bridges near zero frequency, forming a UFO like figure, and this indicates that the field oscillates during certain time intervals and remains inactive during others. $h\ell1C$ has a chirp-like signal that has a black hole with a very energetic frequency of 800Hz; but in most of the plots the frequency remains stable at values close to 200Hz. The D cases are very similar to the black hole ones.
    }
    \label{fig:spectrogram}
\end{figure}

We apply a Fourier transform to our time-domain signals to convert them into the frequency domain. In the spectrograms in Fig. \ref{fig:spectrogram} we plot frequency versus time to analyze the type o spectra shown by the boson star collision. As we can observe, each of the specters for $l=0$ (left panel) and $l=1$ (right panel) has a particular signature, and models $h\ell0A$, $h\ell0B$, and $h\ell0C$ with $l=0$ present the same characteristic in interval time $0.020$\,s to $0.035$ s. The model $hl0D$ with $l=0$ and models $h\ell1C$ and $h\ell1D$ for $l=1$ looks more similar to a chirp obtained for binary black holes. The case being more prominent in model C with $l=1$ showing very similar to the black hole chirp. All models with $\ell=0$ exhibit a spectrum near 300Hz, except $h\ell0D$,which displays a spectrum at 800Hz with a characteristic signal more pronounced than the others. The models $h\ell1A$ and $h\ell1B$ show similar shapes in an interval time of $0.015$ s to $0.035$ s, while $h\ell1C$ has a chirp-like signal as a black hole between $0.015$\,s to $0.035$\,s. $h\ell0D$ and $h\ell1D$ exhibit a similar form, but at different frequencies, this is an indication that the boson stars collapse and they vibrate with a characteristic frequency as the one of black hole they formed as was seeing in Table~\ref{tab:models}. The other five cases presented in the plots, show bridges near zero frequency, forming a UFO like figure; this indicates that the field oscillates, with a constant frequency during certain time intervals and remains inactive during others. We stress the fact that these signals are clearly different from the ones of black hole spectrogramas and, if a signal like that is observed in the LIGO data, could indicate the passage of a gravitational wave produced by the collision of $\ell$-boson stars. 

\begin{figure}[H]
    \centering
    \begin{tabular}{cc}
    \includegraphics[width=0.46\textwidth]{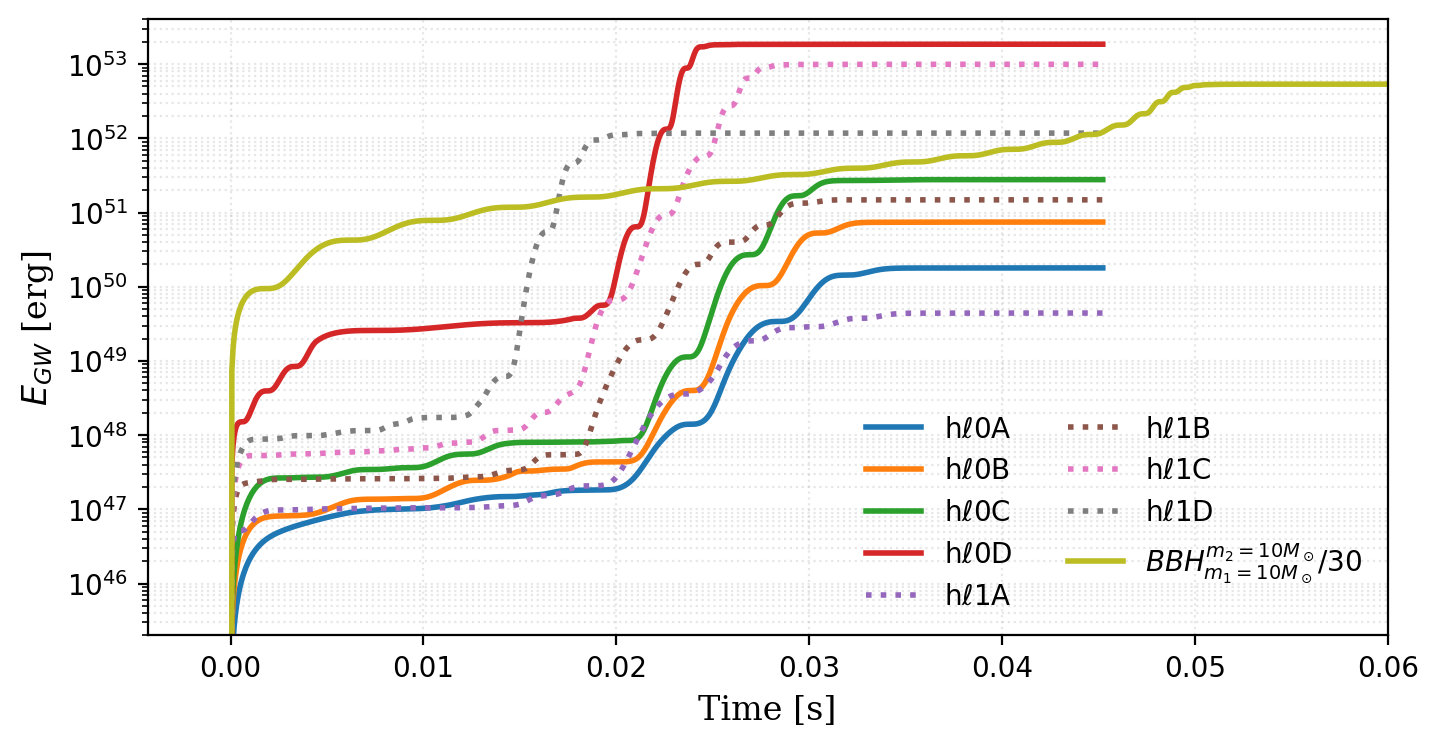}
    & 
    \includegraphics[width=0.45\textwidth]{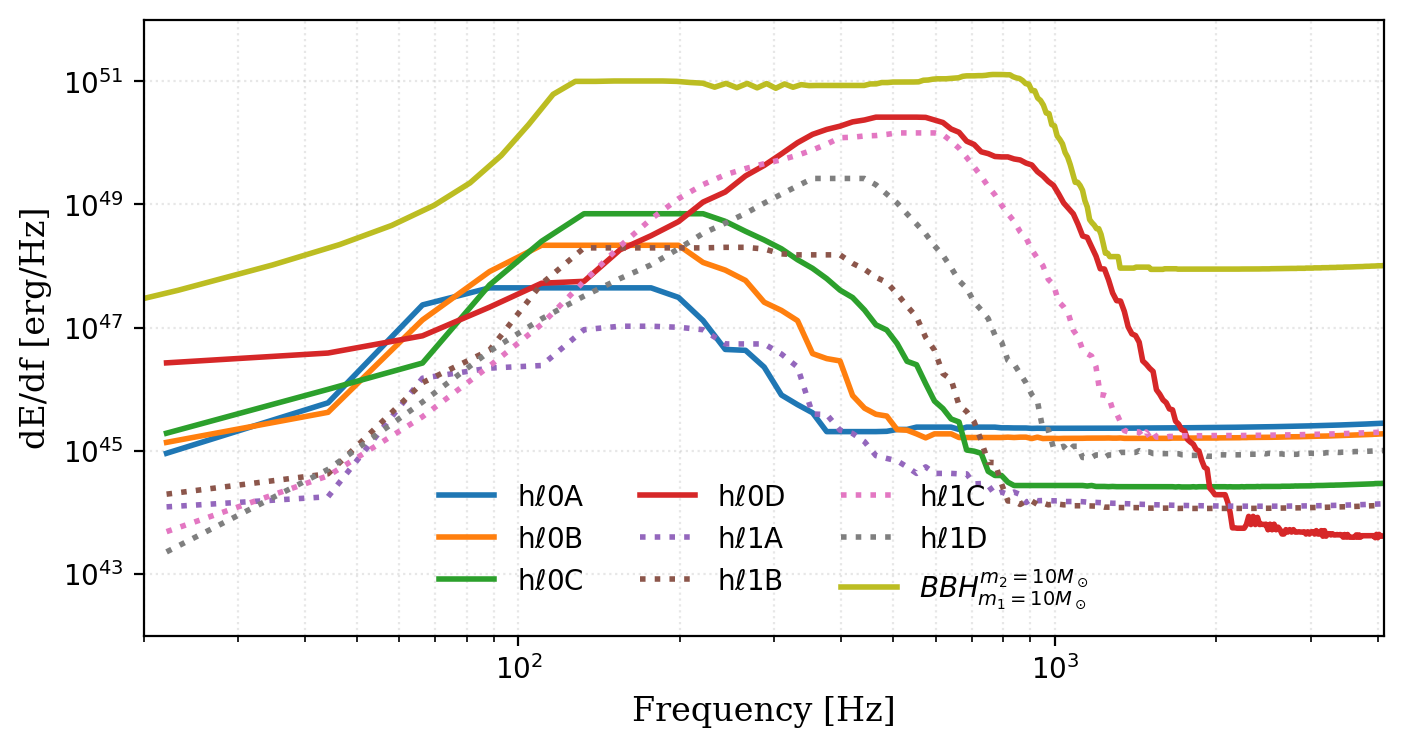}
    \\
    (a) & (b)
    \end{tabular}
    \caption{The gravitational wave signals from boson stars are typically broadband with the majority of the energy at higher frequencies. All the signals indicate a plateau and then a rise around 0.02s, when the boson star merger happens. The model $h\ell0D$ is the more energetic one according to figure (a). We have included the corresponding signal, scaled, of a black hole collision, for comparison. In figure (b) we observe a maximum peak frequency between 100 and 1000 Hz in all models. The quantity $dE_{GW}/df$, which is a function of $f$ and plotted in this figure, is related to the time-integrated radiation energy in the the infinitesimal interval $f$ and $f+df$, $(dE_{GW}/df)\,df$. The obtained spectra of $E_{GW}$ is certainly not monochromatic and resembles (qualitatively) the frequency spectrum of the (quantum) black body radiation.
    }
\label{fig:energy}
\end{figure}

In Fig. \ref{fig:energy}(a) we present the energy versus time behaviour of our signals, and include, for comparison, the one of due to a binary black hole collision. It is straightforward to notice that all the signals have the same behaviour: they present a sharp increase, then plateau and then a climb, roughly with the same slope. The compactness of the stars causes faster stability and the longest slope in energy; we observe that the model $h\ell0D$ has higher energy and $h\ell0A$ less energy, the energy range in the final stage lays between 
$10^{49}$ erg to $10^{53}$ erg. As mentioned above, we have included the corresponding case for the black hole collision, to compare the behaviours; notice that the black hole signal is more energetic and we have scaled it by a factor of $1/30$ to make it fit in the figure; also, we notice the absence of a two stage process, having a sharp increase and then a period of constant augment in the energy until the final value is reached.  

The maximum frequency peak $f_{peak}$ of the power radiation $dE_{GW}(f)/df$ for the waveforms analyzed in this work, are shown in Fig. \ref{fig:energy}(b). Once more, the case $h\ell0A$, presents the most energetic source. Such plots are related to the time-integrated radiation energy in the the infinitesimal interval $f$ and $f+df$, $(dE_{GW}/df)\,df$. The obtained spectra of the energy $E_{GW}$ as a function of the frequency, is certainly not monochromatic and resembles (qualitatively) the frequency spectrum of the (quantum) black body radiation. 

\begin{figure}[H]
    \centering
    \includegraphics[width=0.8\textwidth]{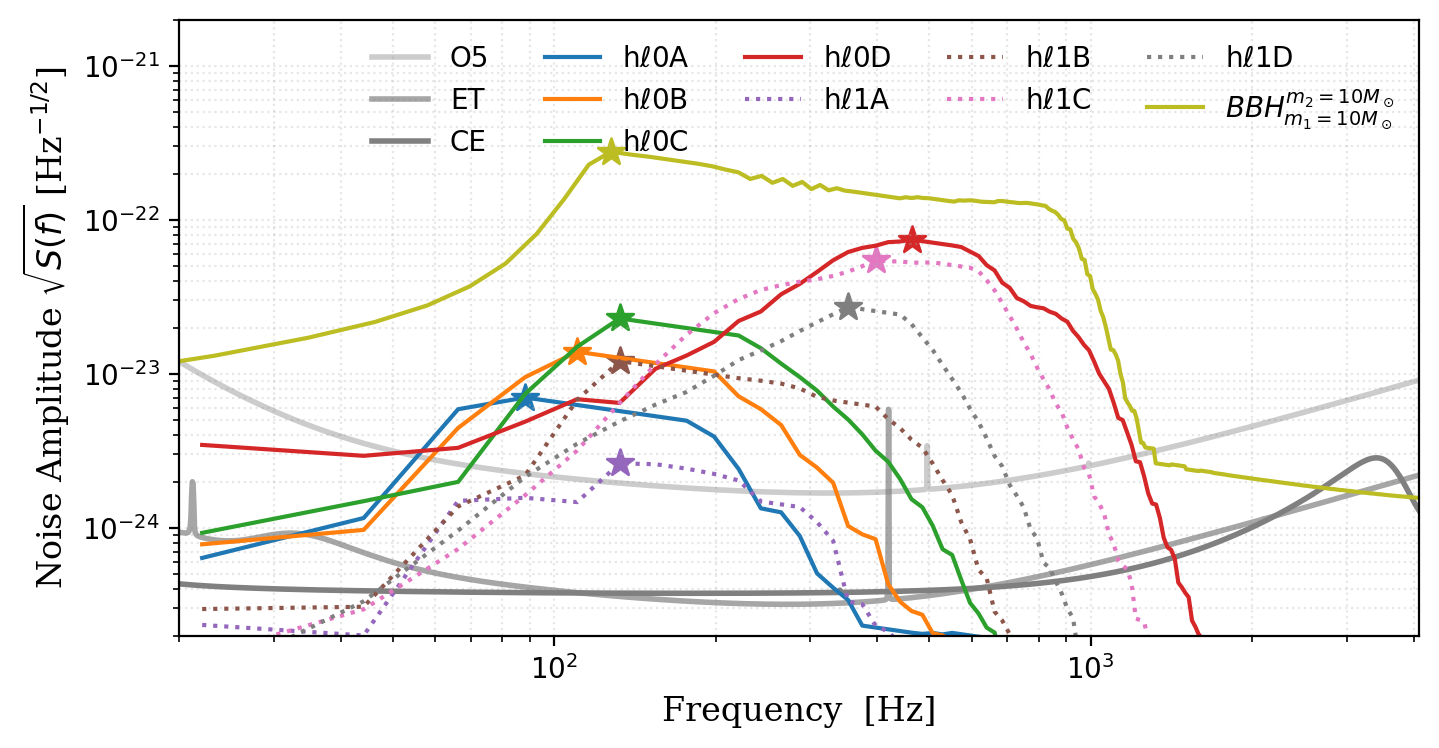}
    \caption{
    Characteristic strain of the $\ell$-boson star GW signals for the case of a scalar field mass of $m_\phi c^2 =  10^{-11}$ eV. The $\ell=0$ boson star signal
    $h\ell0D$ is the strongest and with a wider broadband. The signal of a binary black hole system with $m_1 = m_2 = 10\,M_{\odot}$ and $d=100$ Mpc, yellow line, is shown for comparison.
    For reference, the noise amplitudes of the future Handford and Livingston O5 LIGO, Einstein Telescope and Cosmic Explorer detectors are also shown togheter with all the $h\ell0$ and $h\ell1$ signals.
    }
    \label{fig:energy_all}
\end{figure}

Considering the same scalar field mass for all the signal, allows to compare their detection features.  
We choose a mass of $m_\phi c^2 = 10^{-11}$eV, so that all the signals are within the detection range (this implies that the mass of each binary $\ell$-boson star changes according to Table~\ref{tab:models}). We compare these signals with the noise amplitudes from LIGO’s O5 run at the Hanford and Livingston sites, alongside noise amplitude curves for the Einstein Telescopes and Cosmic Explorer detectors \cite{Hild:2010id}. The sensitivities of these gravitational wave detectors influence their ability to identify features of gravitational waves. In Fig. \ref{fig:energy_all}, the noise amplitude is displayed for all the signals and we have included the binary black hole signal, peaking at around 200 Hz with a strain amplitude of $10^{-21}$, for comparison. The $\ell$-boson star signals exhibit lower noise amplitudes than the black hole signal, ranging from $10^{-22}$ to $10^{-23}$, with frequency peaks between 100 and 1000 Hz. The figure demonstrates that all boson star signals are detectable by all instruments. Among these, signal $h\ell 0$D has the highest peak amplitude, whereas signal $h\ell 1$A is the weakest. The boson signals span a frequency range of 10 to 1100 Hz. Reducing the detection distance leads to an increase in signal amplitude; on the other hand, if the signals are from less massive sources, they shift to higher frequencies.

\subsection{Analysis of the signal \texorpdfstring{$h\ell0D$}{hl} and \texorpdfstring{$h\ell1A$}{hl1A} versus several values of \texorpdfstring{$m_\phi c^2$}{mphi}} 

\begin{figure}[H]
    \centering
    \includegraphics[width=.45\textwidth]{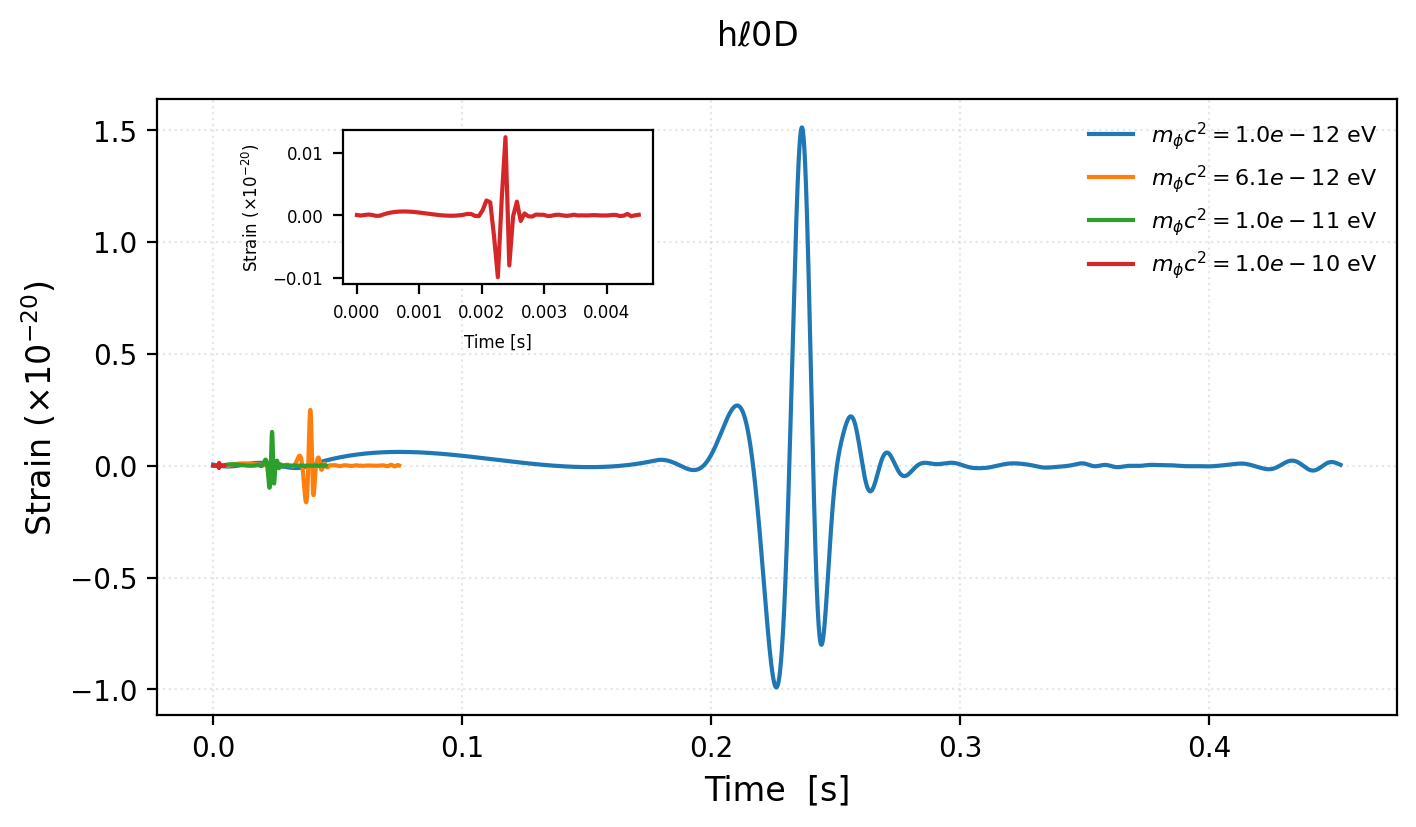}
    \includegraphics[width=.45\textwidth]{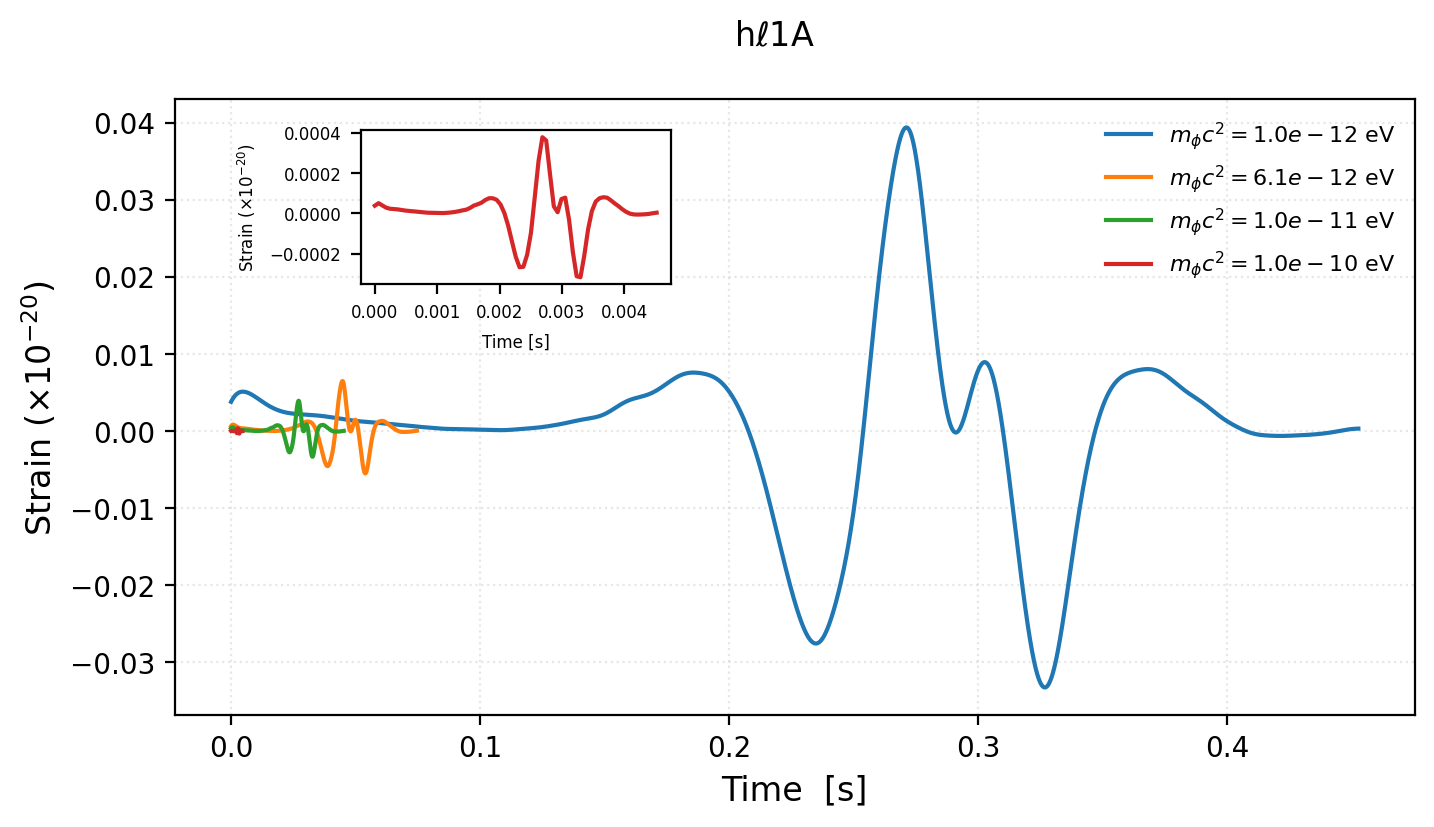}
    \caption{
    The gravitational wave signal model $h\ell0D$ (left panel) and $h\ell1A$ (right panel) are depicted for various values of the scalar field mass $m_\phi c^2$, the signal were generated at a distance of 100 Mpc. The strongest signal occurs for $m_\phi c^2=10^{-12}$ eV. From the figure, it is evident that the amplitude of the $h\ell0D$ model is greater than that of the $h\ell1A$ signal. In the inset we are making and enlargement  signal corresponding to the more masive scalar field.}
    \label{fig:strain_several_mu}
\end{figure}

This section is dedicated to examining the influence of the mass on the scalar field composing the binary $\ell$-boson stars in the model denoted as $h\ell 0$D, due to their pronounced resemblance to the form of binary black hole signals, as well as in the $h\ell 1$A model, which is different, even {\it ad oculum} from a black hole collision signal as can be seen in Fig.~\ref{fig:strain}, where the left panel depicts the waveforms for $h\ell 0$D, while $h\ell 1$A is shown on the right panel. To obtain the corresponding strains, we analyze the parameter interval $10^{-12} < m_\phi c^2 < 10^{-10}$, noting that the mass of the $\ell$-boson star system varies accordingly, as shown in Table~\ref{tab:models}. Fig.~\ref{fig:strain_several_mu} displays waveform signals for all the models described before, where the amplitude and duration of these signals are considerably affected by the mass of the scalar field, while the waveform's shape does not change. In fact, when considering the lightest scenario, the detection time can reach up to 0.3 s, while for the heaviest scenario, it is comparatively one-hundredth of that duration. Notice also that the amplitude in the $h\ell 0$D cases is about $35$ times larger than the corresponding ones of the $h\ell 1$A model.  
An enhanced presence of the scalar field together with a larger scalar field mass, results in higher compactness (refer to Table~\ref{tab:models}) which in turn might be associated to a greater influence of boson particles on the gravitational signal. 

\begin{figure}[H]
    \centering
    \begin{tabular}{cc}
    \includegraphics[width=0.465\textwidth]{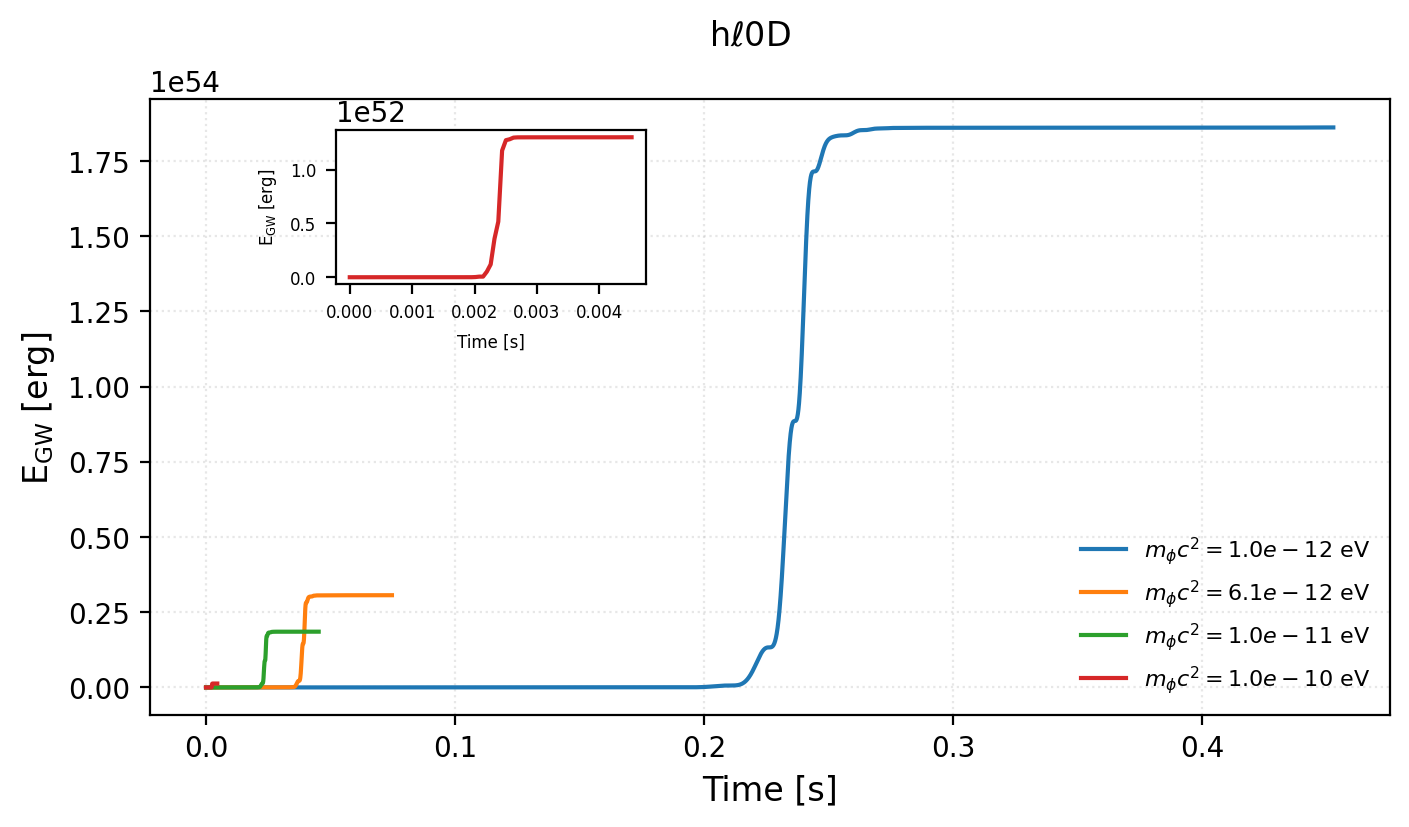}
    & 
    \includegraphics[width=0.45\textwidth]{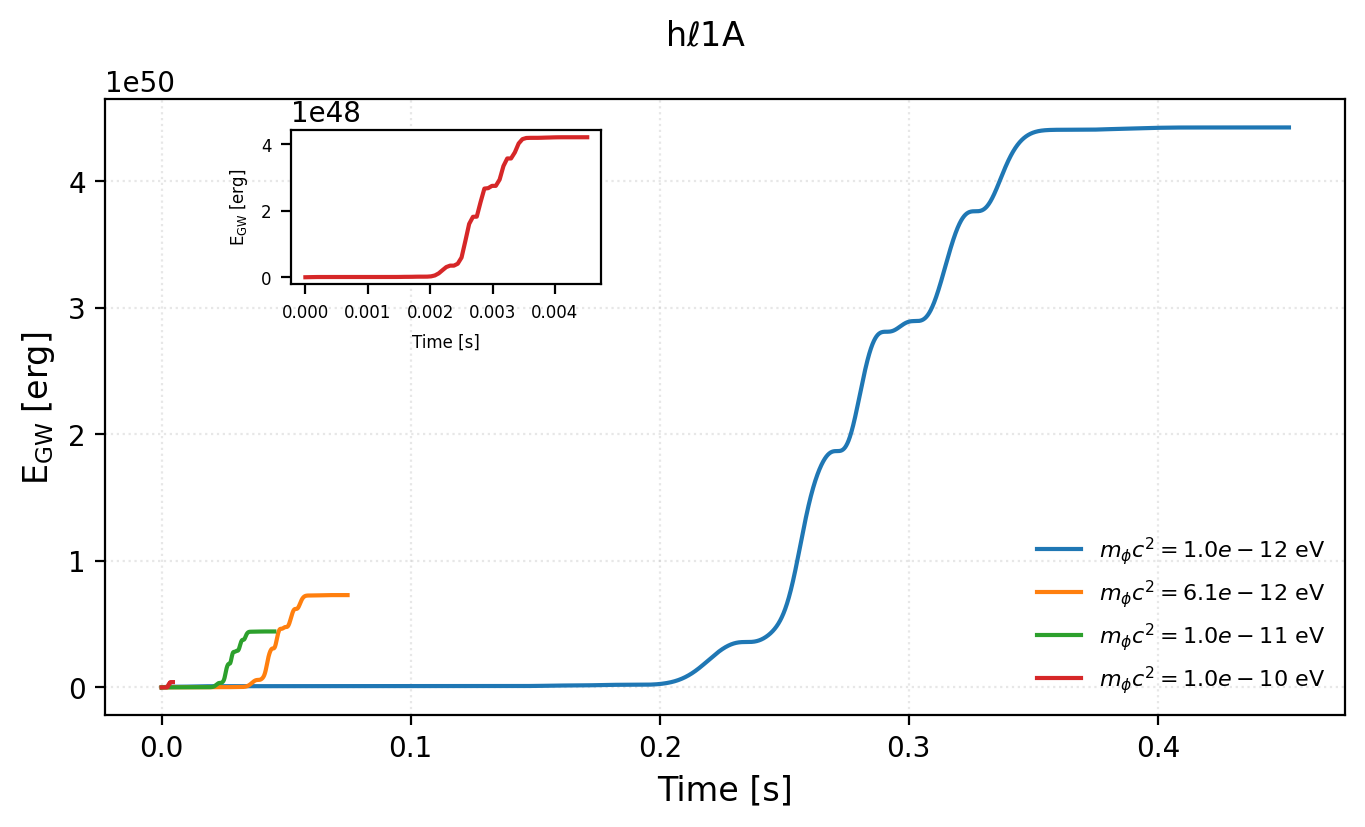}
    \\
    (a) & (b)
    \end{tabular}
    \caption{
    The gravitational energy emissions from $\ell$ boson stars encounters is presented for several values of scalar field mass in both models,  h$h\ell0D$, left, and $h\ell1A$, right. Notice that there is a greater energy linked to stars composed of lighter scalar fields in both  models. For the lighter case, the energy peaks at about 0.2 s. Also, in both models the energy emitted has the same shape for all the scalar field masses, only with a variation in scale. The inset provides an enlarged view of the higher scalar field mass signal.
    }
\label{fig:OneModel_SeveralMu_Energy}
\end{figure}

In Fig. \ref{fig:OneModel_SeveralMu_Energy} we analyze the energy signal over time of all the cases, and we observe that in both sets,  $h\ell 0$D (left panel) and $h\ell 1$A (right panel), the  case corresponding to the lighter scalar field mass displays higher energy than the heavier ones. This is due to the increased compactness of the models with lighter scalar field mass. The models $h\ell 0$D  with mass $m_\phi c^2=10^{-12}\,{\rm eV}$, show a leap around 0.2 s. Additionally, the plot reveals that the emitted energy of the boson star binary system reaches a maximum of $10^{54}$ erg. We also plot the energy for greater values of the scalar field mass, and see that the same shape remains the same but with a smaller scale. Similarly, the energy for the model $h\ell 1$A with mass $m_\phi c^2=10^{-12}\,{\rm eV}$ increases at 0.2 s, and also is the case with the highest value of the emitted energy. 

In Table~\ref{T:variousvalues}
we present the values of the amplitude, hrss, the energy emitted, the maximal frequency, $f_{peak}$, and the signal to noise ratio for O5 LIGO, Einstein Telescope and Cosmic Explorer, for the different values of the scalar field mass $m_\phi c^2$. This table compares to the corresponding values of a binary black hole signal (last right column). As expected, the model h$\ell$0D has similar values to the ones of the black hole collision, even the hrss is larger than the one of the black hole, and all the values decrease inversely to the scalar field mass. Regarding the h$\ell$1A model, all the values are smaller than the corresponding black hole case, even the one with the lightest scalar field mass.

\begin{table}[htpb]
	\centering
	\begin{tabular}{cccccc} 
	\multicolumn{6}{c}{{Model h$\ell$0D}}  \\
	\hline 
	$m_\phi c^2$ [eV] & \,\,\, $1.0\times10^{-12}$ \,\,\, & \,\,\, $6.1\times10^{-12}$ \,\,\, & \,\,\, $1.0\times10^{-11}$\,\,\, & \,\,\, $1.0\times10^{-10}$ \,\,\, & BBH \\

 \hline \hline
 hrss    & 1.4e-21& 9.6e-23 & 4.6e-23& 1.5e-24&  2.0e-22  \\
 Energy [erg] & 1.9e+54& 3.1e+53&  1.9e+53& 1.3e+52&  1.6e+54\\
 $f_{peak}$ [Hz] & 46.4 & 281.1&  463.7 & 4587.5&  127.6   \\
 snr O5  &  595.7 & 78.4 & 36.0 & 0.2 &  154.9 \\ 
 snr ET  & 3214.0 & 414.3 & 173.7& 1.1 & 820.7  \\
 snr CE  & 5164.5 & 358.7 & 164.0& 1.2 &  755.1  \\
	\hline 
\\ 
	\multicolumn{6}{c}{{Model h$\ell$1A}}  \\ 
	\hline   
	$m_\phi c^2$ [eV] & \,\,\, $1.0\times10^{-12}$ \,\,\, & \,\,\, $6.1\times10^{-12}$ \,\,\, & \,\,\, $1.0\times10^{-11}$\,\,\, & \,\,\, $1.0\times10^{-10}$ \,\,\, & BBH \\
 \hline \hline
 hrss    & 7.9e-23 & 5.26e-24 & 2.5e-24& 7.8e-26&  2.0e-22  \\
 Energy [erg] & 4.4e+50 & 7.3e+49&  4.4e+49& 4.2e+48&  1.6e+54\\
 $f_{peak}$ [Hz] & 13.3 & 80.3 &  132.5 & 1310.7 &  127.6   \\
 snr O5  &  5.9 & 3.0 & 1.7 & 0.04 &  154.9 \\ 
 snr ET  & 100.5 & 16.1 & 9.3 & 0.2 & 820.7  \\
 snr CE   & 139.3 & 19.4 & 9.3 & 0.2  &  755.1  \\
	\hline
	\end{tabular}
	\caption{In this table, we present the corresponding values for the strain, hrss, the energy, the maximal frequency $f_{peak}$ and the signal to noise ratio for O5, Einstein Telescope and Cosmic Explorer versus different values of the parameter $m_\phi c^2$ for the models $h\ell0$D and $h\ell1$A. We are including a column with the corresponding values for a binary black holes collisions with masses similar to the ones of the boson starts, just by comparison.}
 \label{T:variousvalues}
\end{table}

\newpage
\section{Conclusions}
\label{conclusions}

We have explored the gravitational wave signals generated by boson stars, which differ from those produced by black hole collisions, and described how these distinct signals would appear in modern gravitational wave detectors.

In particular, we examined gravitational wave signals arising from the plunges of several $\ell$-boson star models. These scalar field self-gravitating objects are {\it bona fide} solutions to the Einstein-Klein-Gordon system, with stable configurations that could plausibly exist in nature. To provide a comprehensive context, we included a brief review of $\ell$-boson stars, detailing their derivation and properties, to emphasize their potential as sources of gravitational waves. Notably, scalar fields are emerging as compelling candidates for describing dark matter, further increasing interest in these configurations.

Since scalar fields are typically assumed to have no direct interaction with baryonic matter, their presence is inferred through their gravitational effects on observable objects. These effects include the emission of gravitational waves during collisions of scalar field configurations, a key focus of this manuscript. After presenting an overview of $\ell$-boson stars, we used results from prior studies to analyze the gravitational wave profiles generated by the head-on collisions (or plunges) of binary $\ell$-boson star systems. We discussed these findings and examined how such profiles might be observed using current gravitational wave detectors.

The gravitational wave profiles emitted during the collision of two $\ell$-boson stars strongly depend on the stars’ compactness, defined by their mass-to-radius ratio. For highly compact stars, the waveforms resemble those produced by black hole collisions. In contrast, less compact stars still generate gravitational waves during the plunge, but their waveforms are distinctly different, with some resembling signals from supernova core bounce events.

A notable feature of the Einstein-Klein-Gordon system is that the scalar field parameter $\mu$ can be absorbed via rescaling of distance and time. This renormalization ensures that the results are independent of $\mu$, allowing the physical characteristics of a given system (such as size and total mass) to be determined once the free parameters are fixed. We described this procedure in detail, providing several illustrative examples.

To highlight the detectability of gravitational waves from $\ell$-boson star collisions, we analyzed four cases each for $\ell=0$ and $\ell=1$ stars. These cases spanned a range of compactness, from highly compact stars producing waveforms similar to binary black hole collisions, to less compact stars generating unique waveforms. Using the rescaling procedure, we ensured that the wave frequencies fell within the detection range of current gravitational wave observatories. Assuming plunge events occur at comparable distances to observed black hole mergers, we determined the signals these waves would generate in detectors. We also derived key properties of these signals, including the hrss, energy, $f_{\text{peak}}$, and snr, comparing them to those from binary black hole collisions to emphasize both similarities and differences.

Remarkably, our results suggest that such signals are indeed detectable. Furthermore, as we can only find what we search for, it is possible these signals are already present in existing data. We hope this work inspires further research in this direction, as it has undoubtedly motivated us to continue exploring these fascinating phenomena.

\acknowledgments
This work was partially supported by 
the National Key R\&D Program of China under grant No.~2022YFC220010, 
the CONACyT Network Project No. 376127 ``Sombras, lentes y ondas gravitatorias generadas por objetos compactos astrof\'\i sicos", and also by the Center for Research and Development in Mathematics and Applications (CIDMA) through the Portuguese Foundation for Science and Technology (FCT - Fundação para a Ciência e a Tecnologia), references UIDB/04106/2020, UIDP/04106/2020.
DN acknowledge the sabbatical support given by the Programa de Apoyos para la Superaci\'on del Personal Acad\'emico de la Direcci\'on General de Asuntos del Personal Acad\'emico de la Universidad Nacional Aut\'onoma de M\'exico, as well as the hospitality and support given by the Departament d’Astronomia i Astrof\'isica, Universitat de Val\`encia in the elaboration of the present work. CM thanks to CONAHCYT, PROSNI-UDG and PRODEP.


\bibliographystyle{unsrt}
\bibliography{aipsamp.bib}


\end{document}